\documentclass[12pt]{article}

\usepackage{graphicx}
\usepackage{epstopdf}
\usepackage[centertags]{amsmath}
\usepackage{amssymb}
\usepackage{amsfonts}
\usepackage[usenames]{color}
\usepackage[
      colorlinks=true,
      linkcolor=blue,
      urlcolor=blue,
      filecolor=blue,
      linkcolor=blue,
      citecolor=blue,
      pdfstartview=FitV,
      pdftitle={},
        pdfauthor={James Lucietti, Mukund Rangamani},
        pdfsubject={Partition sums},
        pdfkeywords={asymptotics, partitions, BPS operators},
        pdfpagemode=None,
        bookmarksopen=true
      ]{hyperref}

\makeatletter
\@addtoreset{equation}{section}

\makeatletter
\renewcommand\section{\@startsection {section}{1}{\z@}%
                                   {-3.5ex \@plus -1ex \@minus -.2ex}%nn
                                   {2.3ex \@plus.2ex}%
                                   {\normalfont\large\bfseries}}
\renewcommand\subsection{\@startsection{subsection}{2}{\z@}%
                                     {-3.25ex\@plus -1ex \@minus -.2ex}%
                                     {1.5ex \@plus .2ex}%
                                     {\normalfont\bfseries}}

\def\baselinestretch{1.2}
\def\sec#1{\S \ref{#1}}

\def\req#1{(\ref{#1})}

\def\eg{{\it e.g.}}

\def\cf{{\it cf.}}
\def\ie{{\it i.e.}}

\def\etc{{\it etc.}}

\def\CH{{\cal H}}
\def\CI{{\cal I}}

\def\CN{{\cal N}}
\def\CO{{\cal O}}
\def\CP{{\cal P}}

\def\CZ{{\cal Z}}

\def\RR{\mathbb{R}}

\def\Sp{{\bf S}}

\def\A5S5{{\rm AdS}_5 \times \S^5}

\def\ord#1{\[ \, #1 \, \]}

\newcommand{\Li}{\,{\rm Li}\,}
\def\be{\begin{equation}}
\def\ee{\end{equation}}
\def\bea{\begin{eqnarray}}
\def\eea{\end{eqnarray}}

\def\ell{l}
\def\ads#1{AdS$_{#1}$}
\def\quarter{\frac{1}{4}}
\def\eighth{\frac{1}{8}}
\def\Re#1{\textrm{Re} \left(#1 \right)}
\def\mZ{{\mathfrak Z}}
\def\pZ{\Xi}
\def\nZ{\CZ}
\def\lZ{\mZ}
\def\ord#1{\CO\left(#1\right)}

\newcommand{\bb}{{\boldsymbol \beta}}
\newcommand{\bn}{{\bf n}}
\newcommand{\bgam}{{\boldsymbol \gamma}}
\newcommand{\bB}{{\boldsymbol b}}

\title{{\bf \Large Asymptotic counting of BPS operators in superconformal field theories}}

\author{\normalsize
James Lucietti\footnote{james.lucietti@durham.ac.uk}  \, and
Mukund Rangamani\footnote{mukund.rangamani@durham.ac.uk} \\[1.2mm]
\small \sl   Centre for Particle Theory \& Department of
Mathematical Sciences,
\\[-1.5mm]
\small \sl Science Laboratories, South Road, Durham DH1 3LE, United Kingdom. \\
}

\begin{document}

\setlength{\baselineskip}{16pt}
\begin{titlepage}
\maketitle
\begin{picture}(0,0)(0,0)
\put(350, 320){DCPT-08/05}
\end{picture}
\vspace{-36pt}

%Abstract
\begin{abstract}
We consider some aspects of counting BPS operators which are annihilated by
two supercharges, in superconformal field theories. For non-zero coupling, the
corresponding multi-variable partition functions can be written in
terms of generating functions for vector partitions or their
weighted generalisations. We derive asymptotics for the density of states for
a wide class of such multi-variable partition functions. We also point out a particular factorisation
property of the finite $N$ partition functions. Finally, we discuss
the concept of a limit curve arising from the large $N$ partition
functions, which is related to the notion of a ``typical state'' and
discuss some implications for the holographic duals.
 \end{abstract}
\thispagestyle{empty}
\setcounter{page}{0}
\end{titlepage}

\renewcommand{\baselinestretch}{1.4}  %looks better
\renewcommand{\thefootnote}{\arabic{footnote}}

%%%%%%%%%%%%%%%%%%%%%%%%%%%%%%%%%%%%%%%%%%%%

%\tableofcontents

%%______________________________________
\section{Introduction}
\label{intro}
%%______________________________________

Understanding the microscopic origin of black hole entropy is an
important problem with the potential to shed light on Quantum
Gravity. String theory has made remarkable progress in this area;
for supersymmetric asymptotically flat black holes in four and five
dimensions we know how to count the microscopic states in terms of
D-brane degrees of freedom
\cite{Strominger:1996sh,Maldacena:1997de}. However, an important
factor which simplifies the counting in all known examples is the
existence of an effective string picture -- the microscopic theory
in the regime of interest can always be modeled by a 1+1 dimensional
CFT with an effective central charge \cite{Strominger:1997eq}. The
degeneracy of states is then simply obtained by applying the Cardy
formula. With the recent advances in constructing supersymmetric
black holes, we now have examples where this effective string model
is insufficient; for example, the supersymmetric \ads{5} black holes
\cite{Gutowski:2004ez,Gutowski:2004yv,Chong:2005da,Chong:2005hr,Kunduri:2006ek}\footnote{See
also \cite{Kunduri:2006uh,Kunduri:2007qy} for some progress towards determining the most general such
black hole via a classification of their near-horizon geometries.} are not described by an effective string model. To
obtain a microscopic derivation of the degrees of freedom for such
black holes, one needs to directly count the states (or operators)
of the dual $\CN = 4$ super Yang-Mills (SYM) theory.

The counting of operators in $\CN =4$ SYM was initiated in
\cite{Sundborg:1999ue,Aharony:2003sx} and later extended to
supersymmetric operators in \cite{Kinney:2005ej}. More generally,
one can consider the problem of counting chiral operators in $\CN
=1$ theories (these are analogous to counting $\eighth$-BPS states
in $\CN =4$ SYM). There is by now a large literature on counting such operators from various perspectives such as Polya counting, giant gravitons, dual giants, the `plethystic program', character formulae, linear sigma models, \etc, -- see
\cite{Romelsberger:2005eg,Biswas:2006tj,Mandal:2006tk,Benvenuti:2006qr,Martelli:2006vh,Bianchi:2006ti,Butti:2006au,Feng:2007ur,Forcella:2007wk,Grant:2007ze,
Dolan:2007rq,Butti:2007jv,Forcella:2007ps,Romelsberger:2007ec,Forcella:2008bb,Dolan:2008qi}.\footnote{One can also consider BPS states in the $D=3$
M2-brane SCFT and the $D=6$ M5-brane SCFT
\cite{Bhattacharyya:2007sa, Bhattacharya:2008zy}; our discussion
incorporates these cases as well.} In $\CN=1$ SCFTs one always has
an R-symmetry, and eigen-operators of this symmetry must have
discrete quantum numbers. The conformal dimension of such operators
must then satisfy $\Delta \geq 3R/2$, with equality if and only if
the operator is BPS (\ie, annihilated by a supercharge). In
addition, such theories will typically possess global symmetries and
therefore operators will be labeled by quantum numbers corresponding
to these (\eg, $\CN=4$ actually has an $SU(4)$ symmetry which
contains $U(1)^3$). Thus generically, BPS operators in such theories
will be labeled by a set of $k$ integers $n_i$, with the conformal
dimension $\Delta$ determined by some combination of them
(corresponding to the R-symmetry). It is then found that in the
large $N$ limit one typically encounters partition functions of the
form
\begin{equation}
\label{part} \lZ_k(\beta_1, \beta_2, \cdots \beta_k)=
\mathop{\prod_{n_i \ge 0}}_{\sum_{i=1}^kn_i>0}\,
\frac{1}{1-e^{-\mathop{\sum_{i=1}^k}\,\beta_i n_i}} \ .
\end{equation}
The $\beta_i$ are some set of chemical potentials conjugate to the
quantum numbers in question $n_i$. Given such a partition function
one would like to extract the density of states, $d_k(n_1,n_2,\cdots
n_k)$ defined by
\begin{equation}
\lZ_k(\beta_1, \beta_2, \cdots \beta_k)= \sum_{n_1,n_2,...,n_k \geq 0}
d_k(n_1,n_2, \cdots n_k) e^{-\mathop{\sum_{i=1}^k}\, \beta_in_i}
\end{equation}
at least in the limit of large $n_i$.

In the $k=1$ case the partition function (\ref{part}) is the
generating function for the number of partitions of an integer and
corresponds to the classic result due to Hardy and
Ramanujan:\footnote{Hardy and Ramanujan in fact obtained a much
more precise formula which was improved upon later by Rademacher
into an exact formula for $p(n)$.}
\be \label{HR} d_1(n)=p(n) \sim \frac{1}{4n \sqrt{3}} \exp\left(\pi
\sqrt{\frac{2}{3}n} \right)\ee
as\footnote{In this paper we use the mathematician's meaning of
$\sim$; that is, $f(n) \sim g(n)$ as $n \to \infty$  is equivalent to
$\lim_{n \to \infty} \frac{f(n)}{g(n)}=1$.} $n \to \infty$.
Similarly, for $k>1$ it turns out that (\ref{part}) is the
generating function for the number of partitions of the ``vector''
$(n_1,n_2,...,n_k)$ (we will define this precisely below).
Curiously, it appears that asymptotics for the number of partitions
of a vector are hard to come by in the literature.
In~\cite{Dolan:2007rq} leading asymptotics for $\log
d_k(n_1,n_2,..,n_k)$ for large $n_i$ (of the same order) were
determined.\footnote{See also \cite{Lin:2005nh} for the leading
exponential behaviour of the $k=2$ partition function encountered in
the plane wave matrix model.} One of the main purposes of this paper
is to improve\footnote{Observe that if $\log d(n) \sim g(n)$ it does
not follow that $d(n) \sim e^{g(n)}$. The converse, though, is of
course true.}  upon this and derive the leading asymptotics for
$d_k(n_1,n_2,...n_k)$ when all the $n_i$ are large (and of the same
order). Thus our results will be direct analogues of (\ref{HR}) for
the number of vector partitions.

A technical objection to determining such
asymptotics would be that such multivariable partition
functions do not seem to have any obvious modular properties,
analogous to the single variable case (although see \cite{Cardy:1991kr} for modular properties of free massless scalars in higher dimensions). Indeed the $k=1$ case due to
Hardy and Ramanujan can be deduced from the modular properties of
the Dedekind function. However, one can also obtain (\ref{HR}) using a method due to Meinardus~\cite{Meinardus:1953fk, Andrews:1998uq}, which does not employ any modular properties, and indeed we will show that
the same method can be used to analyze these multivariable partition
functions.

For $k>1$ the special case $\beta_i=\beta$ of the
above partition function simplifies to
\begin{equation}
\lZ_k(\beta)= \prod_{n=1}^{\infty} \frac{1}{(1-e^{-\beta n})^{a_n}}
\qquad \textrm{where} \qquad  a_n = \;\; ^{n+k-1}C_{k-1}
\label{anpartsum}
\end{equation}
which is a generating function for $c_k(n) \equiv
\sum_{n_1+n_2+...+n_k=n} d_k(n_1,n_2,...,n_k)$. At first sight it
seems that the asymptotics in this special case can be deduced from
the theorem of Meinardus \cite{Meinardus:1953fk}, a direct
generalisation of (\ref{HR}) which applies to generating functions
of the form (\ref{anpartsum}) for some class of sequences $a_n$. The
class of sequences $a_n$ to which the theorem applies is encoded in
the behaviour of the associated Dirichlet series
\begin{equation}
D(s) = \sum_{n=1}^{\infty} \frac{a_n}{n^s} \ .
\label{dirser}
\end{equation}
In particular it is restricted to situations where the Dirichlet
series $D(s)$ has one simple pole. It is straightforward to show
that the Dirichlet series defined by the $a_n$ above
(\ref{anpartsum}) actually has $k$ simple poles. It should be noted
that if one simply requires the leading asymptotic for $\log
c_k(n)$, then one can use a trick employed
in~\cite{Benvenuti:2006qr,Dolan:2007rq, Balasubramanian:2007hu}.
This involves writing (\ref{anpartsum}) as a product of $k$
generating functions whose associated Dirichlet series each have one
simple pole, and observing that the leading asymptotic for $\log
c_k(n)$ comes from the Dirichlet series with the rightmost pole
(this is equivalent to the rightmost pole of (\ref{dirser})).
However, this trick does not seem to generalise to determining the
asymptotic behaviour of $c(n)$. For this we return to the proof of
Meinardus' theorem and consider associated Dirichlet series with
multiple poles. This is not a fundamental obstacle and can be
straightforwardly accounted for; therefore we obtain the leading
asymptotic behaviour of $c_k(n)$ as $n \to \infty$.

In addition we discuss aspects of the finite $N$ partition function. This is based on the fact that generating functions for the finite $N$ partition function have been written down. These partition functions usually take the form of a grand canonical partition function (where $N$ is the particle number) for bosons in a $k$-dimensional harmonic oscillator potential. Combinatorially they can be interpreted in terms of restricted vector partitions, \ie, into partitions with no more than $N$ parts.  By exploiting the combinatorial properties of the partition functions we easily write down explicit expressions for the finite $N$ partition functions and prove an interesting factorization property.

We also discuss how one can use the data in the partition sum to
talk about typical states in the ensemble. The latter while being an
interesting issue in its own right in statistical mechanics, is of
great interest in the holographic context. The typical states that
dominate the ensemble have gross features of black hole geometries
in the following sense: in quantum gravity it is these states that
are expected to have a sensible semi-classical background. One issue
however, is that our discussion will be confined to situations where
we have states preserving two or more supercharges; this class of
states (or their dual operators) in four dimensional superconformal
theories do not include the supersymmetric \ads{5} black holes.
Indeed, the entropy of the ensembles we consider scales as
$N^\alpha$ with $\alpha < 2$; this is insufficient in the AdS/CFT
context to produce sufficient back-reaction to obtain a
semi-classical geometry with a regular event horizon. Nevertheless,
BPS operators preserving higher amounts of supersymmetry provide a
useful toy model to understand aspects of emergent gravity \cite{Berenstein:2005aa}.

We begin with a brief introduction to the class of partition
functions we focus on in \sec{chipfns} and describe how these are
related to vector partitions in \sec{vectorp}. In \sec{finiteN} we
discuss aspects of the partition function at finite $N$ and show how
one can recover explicit formulae exploiting the combinatorial
properties of vector partitions. In \sec{dos} we analyze the density
of states of these partition functions, generalizing the results of
Meinardus. We then address issues regarding the typical states in
ensembles under consideration in \sec{limcr} using the notion of
limit curves for the partition sums. Finally, in \sec{discuss} we
discuss some of the implications our results have for the
understanding of supersymmetric operators in field theories in the
holographic context.

%%______________________________________
\section{BPS cohomology partition functions}
\label{chipfns}
%%______________________________________

The class of field theories we are interested in are superconformal field theories with a minimum of four supercharges,\footnote{We count here just the Poincar\'e  supersymmetries; these theories contain superconformal symmetries as well.}  on $\mathbb{R}^D$ for $3\le D\le 6$. As usual, operators in these theories are related to states on $\mathbb{R} \times \Sp^{D-1}$. These theories have some supersymmetry generators $Q$ and superconformal generators $S$. Standard radial quantization results in the hermiticity conditions $Q^\dagger = S$ (see \cite{Minwalla:1997ka} for a nice discussion of radial quantization in SCFTs). The Hamiltonian $H$ (the dilatation operator), the R-symmetry generators $R_i$  (for a generic $\CN=1$ SCFT there would be just one) and the spins $J_i$ (which come from the rotational symmetry group of $\mathbb{R}^D$), all label the operators of the theory and satisfy (schematically)
\be
\label{SQ}
\{ S, Q \} \sim H-\sum_i R_i -\sum_i J_i \ee
and therefore in a unitary representation one has a BPS bound $H \geq \sum_i R_i +\sum_i J_i$. Operators which saturate this are referred to as BPS. We will call any remaining global symmetries of the theory $G_i$ (these are not present in $\CN=4$ SYM, but generic $\CN=1$ SCFTs may have them). A $Q$-cohomology\footnote{If one is considering a cohomology with respect to more than one $Q$ then it is necessary that the $Q$s commute or anti-commute amongst each other, to ensure the cohomologies are compatible.} consists of operators that are annihilated by some subset of the $Q$ that are not $Q$ exact. This is equivalent to being annihilated by $Q$ and $Q^{\dagger}$ (by analogy with Hodge theory). Therefore from (\ref{SQ}) the $Q$-cohomology and BPS operators are in one to one correspondence, and this will be referred to as the BPS cohomology. Note that the BPS cohomology includes both primaries (annihilated by all $S$) and descendants with the same $R$-charges.

We will be interested in the partition functions over the BPS cohomology of at least two supercharges at weak non-zero coupling, with chemical potentials for each of the $R$-symmetries and global symmetries. The Hamiltonian is determined in terms of the charges as given in \req{SQ}; denoting the Hilbert space of these BPS operators as $\CH_{BPS}$, the partition functions generically  take the form
\begin{equation}
Z = \textrm{Tr}_{\CH_{BPS}} \exp( -r_i\,R_i -g_i\,G_i - \zeta_i\, J_i)
\end{equation}
where $r_i,g_i, \zeta_i$ are some set of chemical potentials conjugate to the symmetry in question. For example, such partition sums have been computed for BPS operators preserving various fractions of supersymmetry in $\CN=4$ SYM in~\cite{Kinney:2005ej}, and for  $\eighth-$BPS states in the M2-brane CFT and $\quarter $-BPS in the M5-brane CFT in~\cite{Bhattacharya:2008zy}, as well as a number of $\CN=1$ SCFTs in $D=4$~\cite{Benvenuti:2006qr,Grant:2007ze} for mesonic operators and in \cite{Butti:2007jv} for baryonic operators. Note that all these calculations assume that counting operators in the classical BPS cohomology gives the same answer as counting in the quantum corrected cohomology.

The $D=4$ case is of particular interest.\footnote{Systematics of the representation theory of the $\CN=2$ and $\CN =4$ superconformal algebras were developed in \cite{Dolan:2002zh}.} The $\CN=4$ result, was actually computed using a trick. In this case $\eighth $-BPS states are chiral operators and thus their cohomology is equivalent to the chiral ring. Properties of the chiral ring could then be used to deduce that the finite $N$ partition function over such operators is governed by an effective multi-dimensional harmonic oscillator model. The concept of a chiral ring in $D=4$ only requires the existence of $\CN=1$ supersymmetry.  Recall that for $\CN =1$ field theories it can be argued \cite{Cachazo:2002ry} that the chiral ring is generated by the gauge invariant combinations of chiral superfields in the theory $\Phi_i$ and the gluino superfield $W^\alpha$, modulo some constraints. The constraints are the F-term constraints $\partial_{\Phi_i} \, W =0$, with $W$ being the superpotential,\footnote{For $\CN=4$ SYM the constraint reads $[\Phi_i,\Phi_j]=0$.} along with $[\Phi_i, W_\alpha\} =0$ and $\{W_\alpha, W_\beta\}  =0$. We refer the reader to \cite{Cachazo:2002ry} for an excellent account of chiral rings in $\CN =1$ supersymmetric field theories. One might expect that the chiral ring can be exploited to write down partition functions for BPS operators in $\CN=1$ SCFTs (as was done in $\CN=4$); this would involve solving the F-term constraint which depending on the superpotential can be a complicated affair. The ``plethystic program'' \cite{Feng:2007ur} and gauged linear sigma model techniques \cite{Grant:2007ze} have been used to deduce partition functions for the chiral ring in a wide variety of $\CN=1$ SCFTs (toric quiver gauge theories), some examples of which we will discuss shortly.

Rather than being concerned with the specific partition functions that occur in given theories, in this paper we will study certain partition functions which can be used as building blocks of the known examples discussed above. There are two basic kinds corresponding to BPS bosonic operators and BPS fermionic operators.

Let us consider the partition function for BPS bosonic operators at finite $N$, which we will denote by $\nZ_k(\bb;N)$ where $\bb=(\beta_1,\cdots \beta_k)$ is a set of $k$ chemical potentials conjugate to $n_i$, the quantum numbers of the various conserved charges in the superconformal theory in question and $N$ is the rank of the gauge group\footnote{For simplicity we will focus on SCFTs whose gauge group is $U(N)$.}. The archetypal example of such a partition function is
\begin{equation}
\pZ_k(\bb ,p) = \prod_{n_1,n_2,\cdots n_k \geq 0}^{\infty} \, \frac{1}{1-p \,\exp(- \bb \cdot \bn)}
\label{zkdef}
\end{equation}
where the infinite product converges if $|p|<1$ and $\Re{\beta_i}>0$, which is actually a generating function for $\nZ_k(\bb;N)$:
\begin{equation}
\pZ_k(\bb,p) = \sum_{N=0}^{\infty} \, \nZ_k(\bb;N) \, p^N . \label{Zpexp}
\end{equation}
Observe that this is the grand canonical partition function for bosons in a $k$-dimensional harmonic oscillator potential, where $p$ is the fugacity (the chemical potential that keeps track of particle number $N$). The partition functions given above correspond for instance to $\frac{1}{2}-$BPS or $\frac{1}{4}-$BPS  states in $\CN=4$ SYM when $k=1,2$  respectively, or to $\frac{1}{8}-$BPS states  states in the M2-brane world-volume theory for $k=4$ and to $\frac{1}{4}-$BPS states in the M5-brane world volume theory  for $k=2$  (the $(2,0)$ SCFT in six dimensions).

Certain BPS cohomologies can include fermionic operators too. The
partition function in this case typically consist of factors coming
from bosonic operators (via products of $\Xi_k(\bb,p)$) as well as
factors in the numerator of the form\footnote{Generically, we will
denote the fermionic analogue to a bosonic quantity by adding a
superscript $f$.}
\be \label{upkdef} \Xi^f_k(\bb,p)=
\prod_{n_1,n_2,...n_k \geq 0 }^{\infty} (1+p\exp(-\bb \cdot \bn )) =
\sum_{N=0}^{\infty} \nZ^f_k(\bb;N) \, p^N \; .\ee
An important  example of
a BPS cohomology which includes fermions is the $\eighth$-BPS cohomology
for $\CN=4$ SYM. It is given by \cite{Kinney:2005ej}
\begin{eqnarray}
\pZ^{(\frac{1}{8})}(x_1,x_2,x_3, \zeta,p) &=& \prod_{l,q,r=0}^\infty \, {\prod_{s=\pm1} \left(1 +p\,  e^{\zeta \,s}\,
x_1^{2l+1} \, x_2^{2q+1} \, x_3^{2r+1} \right)\over \left( 1 -p\,
x_1^{2l} \, x_2^{2q} \, x_3^{2r} \right) \, \left( 1 -p\,
x_1^{2l+2} \, x_2^{2q+2} \, x_3^{2r+2} \right) }\nonumber  \\
&=& \Xi_3(\bb, p)\; \Xi_3(\bb,p\,x_1^2x_2^2x_3^2)\; \prod_{s=\pm 1}
\Xi^f_3(\bb,p\,e^{\zeta\,s}\, x_1x_2x_3) \label{piecesofeight}
\end{eqnarray}
where in the second line we have shown how to rewrite it in terms of our basic partition functions \req{zkdef} and \req{upkdef} defining $x_i^2= e^{-\beta_i}$.

$\CN =1$ superconformal quiver field theories dual to AdS$_5$ $ \times X_5$ with $X_5$ being a Sasaki-Einstein manifold have more complicated partition functions because of the presence of mesonic and baryonic operators (see~\cite{Butti:2007jv} for a recent discussion of the moduli space and \cite{Balasubramanian:2007hu} for an analysis of typical states in such field theories). However, it has been argued that such partition functions can be written as sums over baryon number, with the summands each looking like ``weighted'' versions of the partition function  (\ref{zkdef}) (\ie, each factor in the infinite product is raised to some $\bn$-dependent factor, which also depends on the baryon number).  Hence we will also consider the more general class of partition functions
\be
\label{weight}
\pZ_k^w(\bb,p)= \prod_{n_1,n_2,...,n_k \geq 0} \frac{1}{(1-p\exp(-\bb \cdot \bn))^{w_{\bn}}}
\ee
where $w_{\bn}$ is some sequence labelled by $\bn$ (such that the infinite product converges for $|p|<1$ and $\textrm{Re}( \beta_i) >0)$. For example, the partition function for mesonic operators (zero baryon number) in the quiver field theories dual to AdS$_5 \times Y^{p,q}$ is given by (\ref{weight}) with $k=4$ and $w_{\bn}= \delta_{\bn\cdot{\bf Q},0 }$ where ${\bf Q} = ({p}+{ q},{ p}-{ q},-{ p},-{ p})$ is a charge vector characterising the Sasaki-Einstein manifold $Y^{{ p},{q}}$ \cite{Benvenuti:2006qr,Grant:2007ze} (one can generalize to $L^{abc}$ --  in this case the charge vector is  ${\bf Q} = (a,b,c,-(a+b+c))$).  Of course, the weighted partition functions can be used to build the other partition functions so far discussed; for example if $w_{\bn}=-1$ then $\pZ_k^f(\bb,p)= \pZ_k^w(\bb,-p)$.

We will be dealing with partition functions whose large $N$ limit exists for all values of the chemical potential (\ie, it is $\ord{1}$).\footnote{Certain partition function exhibit large $N$ phase transitions for small chemical potentials. We will not consider such partition functions. We also only consider the large $N$ limit where we keep the chemical potentials $\beta_i$ fixed. As discussed in \cite{Kinney:2005ej}, scaling the chemical potential with $N$ in the large $N$ limit results in an interesting Bose-Einstein condensation.} Indeed the strict large $N$ limit of $\nZ_k(\bb;N)$ exists and we will call it $\lZ_k(\bb)$. The large
$N$ limit can be obtained via the following general trick:
\begin{equation}
\label{largeN} \mZ_k(\bb) = \lim_{p \to 1}\,  (1-p)\, \pZ_k(\bb,p)
\end{equation}
which can be proved as follows. Consider
\begin{equation}
\sum_{N=0}^{\infty}\,  a_N \, p^N \equiv\sum_{N=0}^{\infty}\, \left(\nZ_k(\bb,N)-\lZ_k(\bb)\right) \, p^N= \pZ_k(\bb,p)-\frac{1}{1-p} \, \lZ_k(\bb) .
\label{}
\end{equation}
From the fact that $a_N \to 0$ as $N \to \infty$,  it is easy to show that\footnote{This relies on   Abel's theorem on continuity up to the circle of convergence, see \eg~\cite{Whittaker:1996fk}.} $(1-p)\sum_{N=0}^{\infty}\,  a_N \,p^N \to 0$ as $p \to 1$ and hence the result is established. In the case at hand
\begin{equation}
\lZ_k(\bb) =\prod_{n_1,n_2,\cdots n_k \geq 0, \;  \bn \neq 0}^{\infty}\,  \frac{1}{1- \exp(- \bb \cdot \bn)}
\label{lzkbos}
\end{equation}
If there are also fermionic factors in the partition function, one
simply sets $p=1$ in these, and thus we define $\lZ_k^f(\bb) \equiv \pZ_k^f(\bb,1)$.\footnote{Note that $\lim_{N \to \infty} \nZ_k^f(\bb;N)=0$ using (\ref{largeN}); however such a limit never occurs as fermionic partition functions are always accompanied by a bosonic one which takes care of the factor $1-p$ leading to a finite answer for the large $N$ limit.}

To summarise, there are three basic objects we will focus on; the
grand canonical partition sum $\pZ_k(\bb,p)$, the finite $N$
partition sum $\nZ_k(\bb;N)$ and finally the large $N$ partition sum
$\lZ_k(\bb)$. As explained above we are also interested in the
fermionic generating functions $\Xi^f_k(\bb,p)$ and
$\nZ_k^f(\bb;N)$, however since their analysis is so similar to the
bosonic case we will only indicate the necessary differences at the
end of each section as appropriate. Similar comments apply to the more general weighted partition functions.

One partition function which motivated our analysis is that which
encodes the counting of the chiral ring $\frac{1}{8}-$BPS operators
in $\CN =4$ SYM  discussed above (\ref{piecesofeight}). In
particular, one would like to be able to use the spectral data
contained in the partition sum to construct an effective model
encapsulating the dynamics of these BPS operators. Of greater
interest would be to analyze the partition function of
$\frac{1}{16}-$BPS operators (these are outside the scope of our
analysis as such operators are only preserved by only one supercharge in
our language). In this case one knows the free field theory answer
\cite{Kinney:2005ej}, as well as the count of planar operators \ie ,
graviton states with dimensions of $\CO(1)$,  at strong coupling
\cite{Kinney:2005ej} (using supergraviton representations) and at
non-zero weak coupling \cite{Janik:2007pm} (using the one-loop
dilatation operator). The analysis we undertake is geared towards
applications in the holographic context; in particular,  to address
questions regarding which operators (or states) are expected to be
dual to semi-classical geometries and perhaps to aid in the
construction of  micro-state geometries from the supergravity side
along of the lines of \cite{Lin:2004nb} for $\CN=4$ SYM. We will
revisit these issues in the discussion after a detailed analysis of
the partition sums in the next few sections.

%%______________________________________
\section{Vector partitions}
\label{vectorp}
%%______________________________________

The finite $N$ partition functions $\nZ_k(\bb;N)$ and the large $N$ partition functions $\lZ_k(\bb)$, which we have introduced in the previous section, have a combinatorial interpretation. This is well known to mathematicians, see \eg, \cite{Andrews:1998uq} which we will base our discussion on. Let $P(\bn)$ denote the number of partitions of  $\bn=(n_1,n_2,\cdots n_k)$, an ordered $k$-tuple of non-negative integers not all zero. These are often referred to simply as vector partitions. As for integer partitions one does not account for the order of the parts of the partition;  to take care of this one introduces a concept of an ordering. More precisely one counts the number of distinct ways to write $\bn = \bf{j}^1+\bf{j}^2 + \cdots + \bf{j}^s$ subject to the ordering $\bf{j}^r \geq \bf{j}^{r+1}$, where $\bf{j}^r>\bf{j}^{r+1}$ if and only if $j_i^r>j_{i}^{r+1}$ where $i$ is the least integer such that $j_i^r \neq j_{i}^{r+1}$. Note that this ordering allows us to write the condition $n_1,n_2 , \cdots n_k \geq 0$ and $\bn \neq 0$ simply as $\bn >0$. If the number of parts in the vector partition is restricted to be at most $N$ then we denote the number of such partitions by $P(\bn;N)$. It is then a basic fact that:
\begin{eqnarray}
\lZ_k(\bb) &=& \sum_{\bn \geq 0}  P(\bn) \; e^{-\bb \cdot \bn} \\
\nZ_k(\bb;N) &=& \sum_{\bn\geq 0} P(\bn, N) \;e^{-\bb \cdot \bn}. \label{vectorN}
\end{eqnarray}
For completeness we will give the argument for (\ref{vectorN}) (the arguments for the other cases proceed similarly). Consider the frequency representation of a partition of $\bn$; that is $\bn = \sum_{ {\bf j} >0} a_{{\bf j}} \; {\bf j}$ where $a_{{\bf j}}$ denotes the number of times ${\bf j}$ appears in the partition. Observe that $\sum_{ {\bf j} >0} a_{{\bf j}}$ gives the number of parts in the partition. It is then clear that:
\bea
&&\sum_{N=0}^{\infty} \sum_{\bn\geq 0} \, P(\bn, N) \;e^{-\bb \cdot \bn}\,p^N = \sum_{a=0}^{\infty}\sum_{ \{ a_{{\bf j}} |\;  {\bf j}\, >\, 0 \}} \exp( - \sum_{ {\bf j >0} }\bb \cdot {\bf j} \; a_{{\bf{ j}}} ) \;p^{a+ \sum_{ {\bf j >0} }\, a_{{\bf{ j}}}} \nonumber \\
&&\qquad = \frac{1}{1-p}\prod_{ {\bf j}>0} \sum_{a_{{\bf j}}=0}^{\infty} \; \left(p\,e^{-\bb \cdot {\bf j}}\right)^{a_{{\bf j}}} = \frac{1}{1-p} \;\prod_{ {\bf j}>0} \frac{1}{1-pe^{-\bb \cdot {\bf j}}} \nonumber
\eea
which establishes the result. Note that in the first line each sum over $a_{{\bf j}}$ is from $0$ to $\infty$ and $a= N- \sum_{ {\bf j >0} } a_{{\bf{ j}}}$ (recall we are looking at partitions whose number cannot exceed $N$), and in the second line the various geometric sums have been performed.

We will be interested in the asymptotic density of states for these partition functions, and from the above this can be seen to be equivalent to the asymptotic number of vector partitions.

The fermionic partition functions also have a combinatorical interpretation in terms of vector partitions. Let $Q(\bn)$ be the number of partitions of $\bn$ (as defined above) into  distinct parts, where $(0,0,...,0)$ may be a part. Similarly denote the number of vector partitions of $\bn$ into $N$ distinct parts (including ${\bf 0}$) by $Q(\bn;N)$. It then turns out that the generating functions for these are given by our fermionic generating functions
\begin{eqnarray}
\lZ_k^f(\bb) &=& \sum_{\bn \geq 0}  Q(\bn) \; e^{-\bb \cdot \bn} \\
\nZ_k^f(\bb;N) &=& \sum_{\bn\geq 0} Q(\bn, N) \;e^{-\bb \cdot \bn}.
\end{eqnarray}
%

%~~~~~~~~~~~~~~~~~~~~~~~~~~~~~~~~~~~~~~~~~
\section{Finite $N$ generating functions}
\label{finiteN}
%~~~~~~~~~~~~~~~~~~~~~~~~~~~~~~~~~~~~~~~~~

We will now explore the finite $N$ partition sums in some detail. Our aim is to utilize some of the combinatorial properties of these partitions to give an algorithmic method to compute these from the grand-canonical partition function.  The motivation behind undertaking this exercise is to learn about the operators that are present in the finite $N$ theory. Of course, this is relevant in the context of holography only when we are interested in quantum effects,\footnote{Recall that in the AdS/CFT correspondence, string effects are suppressed by $g_s \sim \frac{1}{N}$.} but the analysis reveals interesting results which should enable one to better understand the set of BPS operators as we will discuss.

We first note that the $k=1$ case is well known:
\begin{equation}
\label{koneN}
\nZ_1(\beta; N)= \prod_{n=1}^N \frac{1}{1-e^{-n\beta}}
\end{equation}
which can be deduced by a variety of methods (for instance, directly using the well known free fermion description of the system or using the q-binomial theorem \cite{Dolan:2007rq}).

We may now exploit a useful trick in extracting the finite $N$ generating function from $\pZ_k(\bb,p)$. Note that the logarithm of the grand canonical partition sum  admits a simple Taylor series in $p$
\begin{equation}
\log\pZ_k(\bb,p) = \sum_{m \geq 1 , \bn \geq 0}\frac{ p^m \,e^{-m \,\bb \cdot \bn}}{m} = \sum_{m=1} \frac{p^m}{m} \; \left(\prod_{i=1}^k\, (1-e^{-m\,\beta_i}) \right)^{-1}.
\label{logxi}
\end{equation}
A general result is that for any function of $G(p)$ such that
\begin{equation}
\log G(p)= \sum_{m=1} \frac{g_m\, p^m}{m!}
\end{equation}
then
\begin{equation}
G(p)= \sum_{m=0} \frac{Y_m \,p^m}{m!}
\end{equation}
where $Y_m=Y_m(g_1,g_2, \cdots g_m)$ are the Bell polynomials \cite{Andrews:1998uq}. They\footnote{Not surprisingly, the Bell polynomials are closely related to the notion of the plethystic exponential introduced in \cite{Benvenuti:2006qr,Feng:2007ur}.} can be defined by
\begin{equation}
Y_m(y_1,y_2,\cdots y_m) = e^{-y} \frac{d^m e^y}{ dt^m} \qquad \textrm{where} \qquad y_i= \frac{d^iy}{dt^i}.
\label{bellpdef}
\end{equation}
The first few are:
\begin{equation}
Y_0=1, \qquad Y_1(g_1)=g_1, \qquad Y_2(g_1,g_2)= g_1^2+g_2, \qquad Y_3(g_1,g_2,g_3)= g_1^3+3\,g_1\,g_2+g_3
\end{equation}
 It can be shown that they satisfy the following recurrence relation:
\be
\label{recursion}
Y_{n+1}(g_1,g_2, \cdots, g_{n+1})= \sum_{m=0}^n \, {}^nC_m\, Y_{n-m}(g_1, \cdots g_{n-m})\, g_{m+1}.
\ee
Thus, in the case at hand $G(p)=\pZ_k(\bb,p)$ and hence, from equation (\ref{logxi}) we can read off
\begin{equation}
g_m(\bb)= (m-1)! \prod_{i=1}^k\; \frac{1}{1-e^{-m\beta_i}}
\label{gmdef}
\end{equation}
 which gives the following expression for the finite $N$ partition function
\begin{equation}
\nZ_k(\bb;N)= \frac{1}{N!}\,  Y_N(g_1(\bb),g_2(\bb), \cdots, g_N(\bb) )\label{finNkpart}
\end{equation}
which is a result known in the mathematical literature \cite{Andrews:1998uq}; this provides a generalisation of (\ref{koneN}) for $k>1$. For low values of $N$ this is easily computed using \req{bellpdef} or the recursion relation (\ref{recursion}).

Similarly, one can work out the finite $N$ partition function for fermionic and weighted vector partitions. The answers in both cases are given by (\ref{finNkpart}) with $g_m(\bb)$ replaced by one of the following: for the fermionic partition function $\nZ_k^f(\bb;N)$
\be
g^f_m(\bb) = (-1)^{m+1} g_m(\bb)
\ee
whereas for the weighted partition function $\nZ^w_k(\bb;N)$
\be
g^w_m(\bb)= (m-1)!\sum_{\bn \geq 0} w_{\bn}\, e^{-m \; \bb \cdot \bn}.
\ee
%

%~~~~~~~~~~~~~~~~~~~~~~~~~~~~~~~~~~~~~~~~~
\subsection{Factorization property of finite $N$ generating functions}
\label{factorfN}
%~~~~~~~~~~~~~~~~~~~~~~~~~~~~~~~~~~~~~~~~~

The Bell polynomials are very useful to infer certain properties of the finite $N$ partition functions directly. They allow one to prove the following

\noindent
{\bf Factorisation property:} The partition function $\nZ_k(\bb;N)$ can be written as
\begin{equation}
\nZ_k(\bb; N)=\left( \prod_{i=1}^k \nZ_1(\beta_i; N) \right) \; \CP_k(\bb;N)
\label{factzkN}
\end{equation}
where $\CP_k(\bb;N)$ is a symmetric polynomial in $x_i = e^{-\beta_i}$ of total degree at most $\frac{1}{2}\, k \, N\,(N-1)$. In particular, when $k$ is even
\begin{equation}
 \CP_k(\bb;N)= \left( \prod_{i=1}^k x_i
\right)^{\frac{1}{2}N(N-1)} + \cdots +1 \label{orderP}
\end{equation}
where the $\cdots$ stand for the non-constant terms of the polynomial of lower
order.\footnote{In fact this result is known in the mathematics literature \cite{Gordon:1963cs}, a fact we were not initially aware of; we therefore provide our own proof of this result. Further, \cite{Gordon:1963cs} also proves that the coefficients of the polynomial are non-negative. }

\noindent
{\bf Proof:} We first note that (\ref{koneN}) implies
\begin{equation}
\prod_{i=1}^k \nZ_1(\beta_i; N) =\prod_{m=1}^N \frac{g_m(\bb)}{(m-1)!}.
\end{equation}
Then the recursion relation (\ref{recursion}) allows us to deduce one for $\CP_k(\bb;N)$:
\begin{equation}
\label{Prec}
\CP_k(\bb;N+1)= \frac{1}{g_1\,g_2 \cdots \,g_{N+1}} \sum_{m=0}^N a_{m,N} \; g_1\,g_2 \cdots \,g_m \; g_{N-m+1} \;  \CP_k(\bb;m)
\end{equation}
where $a_{m,N}= \frac{1}{(N+1)(N-m)!} \,\prod_{n=m}^{N}\,n!$. One can use this to prove that $\CP_k(\bb;N)$ is a polynomial by induction. Assume that $\CP_k(\bb;m)$ is a polynomial for $m \leq N$ and observe the base case $\CP_k(\bb;1)=1$. It is clear that all the terms in (\ref{Prec}) for which $N-m+1>m$ are polynomial in $x_i$ since one can cancel each $g_n$ in the numerator with one in the denominator, thus leaving only a product of some of the $g_n$ in the denominator (recall $1/g_n$ are polynomials in $x_i$). At first glance the terms for which $N-m+1 \leq m$ are not obviously polynomial, as one will have a factor of $g_{N-m+1}^2$ in the numerator but only a $g_{N-m+1}$ in the denominator. For these terms argue as follows. Cancelling $g_1\,g_2 \cdots \, g_m$  in the numerator with the corresponding factors in the denominator will leave
\begin{equation}
\frac{g_{N-m+1}}{g_{m+1}\,g_{m+2} \,\cdots \,g_{N+1}}.
\end{equation}
Note that $m+1, m+2,\cdots, N+1$ is a sequence on $N-m+1$ consecutive integers and thus one of these must be divisible by $N-m+1$. Lets  call this particular integer $M$ so that $M=q(N-m+1)$ for some integer $q$. Then since
\begin{equation}
\frac{g_{N-m+1}}{g_{M}}= \frac{(N-m)!}{(M-1)!}\prod_{i=1}^k \frac{1-x_i^M}{1-x_i^{N-m+1}} = \frac{(N-m)!}{(M-1)!}\prod_{i=1}^k\left(1+ x_i^{N-m+1}+\cdots+ (x_i^{N-m+1})^{q-1} \right)
\end{equation}
is a polynomial we have proved that all the terms in (\ref{Prec})
with $N-m+1 \leq m$ are also polynomial. Therefore we deduce that
$\CP_k(\bb;N+1)$ is polynomial completing the proof by induction.

One can use the above argument to extract the order of the
polynomial $\CP_k(\bb,N)$. This depends on whether $k$ is even or
odd. In particular, note that:
\begin{equation}
 \frac{g_{N-m+1}}{g_{m+1}g_{m+2} \cdots g_{N+1}} = \frac{(N-m)!}{\prod_{n=m}^N n!}\left[
(-1)^{k(N-m)}\left(\prod_{i=1}^k x_i \right)^{ \frac{1}{2}(N+m)(N-m+1)} + \cdots +1 \right]
\label{gexp}
\end{equation}
(which holds for all $0 \leq m \leq N$). First consider when $k$ is even. We will prove (\ref{orderP}) by induction.
The $N=1$ case is trivial. Now assume, for induction, that
(\ref{orderP}) holds for $\CP_k(\bb,m)$ for all $m \leq N$. Using (\ref{gexp}),
(\ref{Prec}) and the induction step gives
\bea \CP_k(\bb, N+1)
&=& \frac{1}{N+1} \sum_{m=0}^N \left[\left(\prod_{i=1}^k x_i
\right)^{ \frac{1}{2}(N+m)(N-m+1)+\frac{1}{2}m(m-1)} + \cdots +1
\right]
\nonumber \\ &=& \frac{1}{N+1} \sum_{m=0}^N
\left[\left(\prod_{i=1}^k x_i \right)^{ \frac{1}{2}N(N+1)} + \cdots
+1 \right]  \nonumber \\ 
&=& \left(\prod_{i=1}^k x_i \right)^{
\frac{1}{2}N(N+1)} + \cdots +1 \eea
 and thus we learn that
(\ref{orderP}) also holds for $\CP_k(\bb,N+1)$. Hence by induction
this proves (\ref{orderP}) for all $N$.

The situation with $k$ odd is tricker, since the leading order term in \req{orderP} tends to get cancelled. While the precise order seems to depend on particular values of $k$ and $N$ it remains true that $\CP_k(\bb; N)$ grows at most as fast as
$\left(x_i\right)^{\frac{1}{2}\, N\,(N-1)}$ in each of the $x_i$.

\noindent
{\bf Remarks:} The above result has an interesting implication. The partition function $\nZ_k(\bb;N)$ is that of $N$-bosons in a $k$-dimensional harmonic oscillator. The factorisation theorem we have proved shows it is equal to the product of $k$ distinct partition functions of $N$-bosons in a $1$-dimensional oscillator $\nZ_1(\beta_i;N)$, times a partition function whose Hamiltonian has a spectrum which is bounded above and below which presumably provides the interaction term between the $k$ systems.  We discuss some implications of this result in \sec{discuss}.

\paragraph{Fermions:} The partition function $\nZ_k^f(\bb;N)$ can be written as
\begin{equation}
\nZ_k^f(\bb; N)=\left( \prod_{i=1}^k \nZ_1(\beta_i; N) \right) \; \CP_k^f(\bb;N)
\label{factzkNf}
\end{equation}
where $\CP_k^f(\bb;N)$ is a symmetric polynomial in $x_i = e^{-\beta_i}$ of total degree at most $\frac{1}{2}\, k \, N\,(N-1)$ with no constant term (for $N>1$). In particular, when $k$ is odd
\begin{eqnarray}
\CP^f_k(\bb,1)&=& 1, \\
 \CP_k^f(\bb;N) &=& \left( \prod_{i=1}^k x_i
\right)^{\frac{1}{2}N(N-1)} + \cdots +0, \qquad N \geq 2\label{orderPf}
\end{eqnarray}
where the $\cdots$ stand for the non-constant terms of the polynomial of lower
order (so this polynomial does not have a constant term for $N \geq 2$).

To prove this one can use the same technique as we employed in the bosonic case, which we will only sketch. We find that $\CP^f_k(\bb;N)$ satisfies the recursion relation (\ref{Prec}) with $a_{N,m}$ replaced by $a^f_{N,m}=(-1)^{N-m}a_{N,m}$, and the proof that $\CP^f_k(\bb;N)$ is a polynomial goes through in the same way. To extract the order of the polynomial one uses (\ref{gexp}), which for odd $k$ will contribute a factor of $(-1)^{N-m}$ to the leading order terms and will cancel with that in $a^f_{N,m}$, thus giving the same leading order answer as the bosonic case. However, the constant terms will cancel due to the alternating sign in $a^f_{N,m}$.

Observe that the $k=1$ case is simply
\be
\nZ_1^f(\bb;N)= e^{-\frac{1}{2}N(N-1)\beta} \prod_{n=1}^N \frac{1}{1-e^{-n\beta}}
\ee
a result which is easily derived by other means (\eg, one can use a method used to derive the $q$-binomial theorem). This result makes sense as $\frac{1}{2}N(N-1)$ is the ground state energy for $N$ fermions in a harmonic oscillator potential. The fact that the constant term in $\CP_k^f(\bb;N)$ (for $N>1$) is absent makes sense as this system has fermionic statistics.

%~~~~~~~~~~~~~~~~~~~~~~~~~~~~~~~~~~~~~~~~~~~~~~~~~
\section{Asymptotic density of states}
\label{dos}
%~~~~~~~~~~~~~~~~~~~~~~~~~~~~~~~~~~~~~~~~~~~~~~~

Consider the partition functions $\nZ_k(\bb;N)$. From a statistical mechanics point of view these encode the degeneracy of states carrying charges $\bn$, which we will denote by $d_{k,N}(\bn)$. Therefore from the combinatorial interpretation of this partition function (\ref{vectorN}) we immediately see that the density of states is given by $d_{k,N}(\bn)=P(\bn, N)$. Thus\be
\label{degdef} \nZ_k(\bb;N)= \sum_{\bn \geq 0} d_{k,N}(\bn)e^{-\bn
\cdot \bb} \ee which can be inverted to give
\begin{equation}
d_{k,N}(\bn) = \prod_{i=1}^k\int_{b_i-\pi i }^{b_i+\pi i} \, \frac{d\beta_i}{2\pi i} \;\nZ_k(\bb;N) \, e^{\bn \cdot \bb}
\label{dosN}
\end{equation}
 for some arbitrary $b_i \in \RR_+$. We will also define $d_k(\bn)= \lim_{N \to \infty} d_{k,N}(\bn)$. We will be interested in the asymptotic behaviour of $d_{k,N}(\bn)$ and $d_{k}(\bn)$ as $n_i \to \infty$. Of course, for $k>1$ there is more than one way of doing this. We will focus on the case where we send all the $n_i \to \infty$ at the same rate. Other cases, such as keeping some subset of the $n_i$ fixed, are more complicated, and we will outline how these can be dealt with at the end.

The integral for $d_{k,N}(\bn)$ at large $\bn$ can be calculated using the method of steepest descent, which requires choosing the contour (and thus $b_i$) such that we pass through the dominant\footnote{We will only encounter cases where there is a unique dominant saddle point.} saddle point in the direction of steepest descent. Note that choosing the dominant saddle means that contributions from any other saddle points will be exponentially surpressed. If we let $b_i \to 0^+$ we expect this integral to be dominated by the $\beta_i \to 0^++i0$ region of integration, since the singularities of $\nZ_k(\bb;N)$ come from those of $g_m(\bb)$ defined in \req{gmdef};  the strongest of these occurs at $\beta_i=0$. Since we are sending the $n_i \to \infty$ at the same rate, we expect that will will need to send the $\beta_i \to 0^+$ at the same rate. Indeed from (\ref{finNkpart}) one can easily deduce that in this case the leading order behaviour is
\begin{equation}
\label{finiteNasymp}
\nZ_k(\bb;N) \sim \frac{1}{N! \prod_{i=1}^k \beta_i^N}.
\end{equation}
The saddle point is located at an extremum of the exponent of the integrand:
\begin{equation}
n_i + \frac{ \partial}{\partial \beta_i} \log \nZ_k(\bb;N) =0
\end{equation}
and thus, using (\ref{finiteNasymp}), one gets $\beta_i \sim N/n_i$. This shows that sending the $\beta_i \to 0^+$ at the same rate is consistent with sending the $n_i \to \infty$ at the same rate.

We will now present a systematic analysis of the asymptotic behaviour of the density of states by generalizing the classic results of Hardy-Ramanujan and Meinardus~\cite{Meinardus:1953fk}.\footnote{See \cite{Actor:1994kx} and \cite{Granovsky:2007lq} for other generalisations.} As a warm up we start on the single chemical potential case obtained by setting $\beta_i = \beta\;\; \forall \; i$. We then discuss the general case of unequal chemical potentials. We will present the results for the large $N$ partition functions $\lZ_k(\bb)$ and indicate some generalisations at the end.

%____________________________________
\subsection{Equal chemical potentials: Meinardus generalized}
\label{eqchemp}
%_____________________________________

When the chemical potentials coincide, the large $N$ partition sums of interest are as in \req{anpartsum}. These are similar to the weighted partitions considered by Meinardus \cite{Meinardus:1953fk,Actor:1994kx} with an important difference. To see this let us start with the Mellin representation of the partition function (\ref{anpartsum})
\begin{equation}
\begin{split}
\log \lZ_k(\beta) & = -\sum_{n=1}^\infty \, a_n \, \log \left( 1-e^{-n \, \beta}\right) = \sum_{n=1}^\infty \, \sum_{m=1}^{\infty} \, \frac{1}{m}\, a_n \, e^{- m \, n \, \beta} \\
&= \frac{1}{2\pi i}\, \int_{\gamma-i\infty}^{\gamma+i\infty} \, ds \,  \Gamma(s) \, \zeta(s+1) \,D(s) \, \beta^{-s}.
\label{logZdir}
\end{split}
\end{equation}
The first line is standard and in writing the second line we have used the Mellin representation of the exponential
\begin{equation}
e^{-x} = \frac{1}{2\pi i}\, \int_{\gamma-i\infty}^{\gamma+i\infty} \, ds \,  x^{-s} \, \Gamma(s)
\label{invMellin}
\end{equation}
and re-summed the series in $m$ to obtain the  Riemann zeta sum $\zeta(s+1)$. Finally,  the summation over $n$ leads to the Dirichlet series, $D(s)$, as defined in \req{dirser}. This is the point of departure from Meinardus; as we will see the Dirichlet series of interest have $k>1$ simple poles.\footnote{This occurs because the $a_n$ of interest \req{anpartsum} are polynomials in $n$ of order $k-1$, and thus $D(s)$ is a linear combination of $k$ zeta functions (each of which have a simple pole) with shifted arguments} These don't seem to have been fully considered in the literature.\footnote{See however \cite{Actor:1994kx} for discussions of spectra in toroidally compactified theories, where zeta-sums with multiple poles arise.} One reason to ignore the subtlety is that generically the dominant contribution to the density of states comes from the rightmost pole (a fact used in the recent analysis of  \cite{Balasubramanian:2007hu}). We will now present the effects of having the $k$ simple poles, which is quite simple to implement algorithmically.

So far we have not specified the precise contour of integration in \req{logZdir}. For convergence of the integral we need the contour to lie to the right of any singularity arising from the Dirichlet series (or the zeta function). This requires that  $\gamma>k$ since the defining expression for $D(s)$ converges for $\Re{s} >k$.

To proceed it is useful to obtain an integral expression for the Dirichlet series
\begin{equation}
D(s)= \frac{1}{\Gamma(s)} \, \int_0^{\infty} \, dt \, t^{s-1} \, g(t) \ ,
\qquad
g(t)= \sum_{n=1}^{\infty} a_n t^n = \frac{1}{(1-e^{-t})^k}-1
\label{dintrep}
\end{equation}
which allows us to  obtain an expression for $D(s)$ valid in the whole complex plane
\begin{equation}
D(s) = -\frac{\Gamma(1-s)}{2\pi i}\, \int_{\infty}^{(0^+)} \, dt \, (-t)^{s-1}\,  g(t) 
\label{dhankel}
\end{equation}
where we have used a standard notation for the Hankel contour.\footnote{The Hankel contour starts just above the real axis at $+\infty$, encircles the origin counter-clockwise, and ends up just below the real axis at $+\infty$.}  To obtain \req{dhankel} we used a standard trick to the one employed in the analytic continuation of the Gamma-function, or indeed the Riemann zeta function \cite{Whittaker:1996fk}. The integral representation \req{dhankel} allows us to deduce that $D(s)$ is meromorphic with simple poles at $s=1,2, \cdots, k$. Let the residues of $D(s)$ at  these poles be $A_j$ where $j=1,2, \cdots, k$. One can show that for the case at hand
\be
A_j=\textrm{Res}_{s=j} D(s)= \frac{(-1)^{j+k}}{(j-1)!} \sum_{n_1,...,n_k \geq 0} \delta_{n_1+n_2+ ... +n_k,k-j} \prod_{i=1}^k \frac{B_{n_i}}{n_i !}
\ee
where $B_n$ are the Bernoulli numbers, and thus note in particular that $A_k=1/(k-1)!$. However, we can proceed with the calculation for general $A_j$.

We now return to \req{logZdir}, where we shift the contour at $\Re{s} = \gamma$ to the left, across the poles of the integrand, to  $\Re{s} = -\epsilon$ with $0<\epsilon<1$. Therefore $\log \lZ_k$ receives contributions from the poles of the integrand in the region $-\epsilon < \Re{s} <\gamma$ (which include those of $D(s)$) and we find
\begin{equation}
\lZ_k(\beta) =\exp\left( \sum_{j=1}^k \; \frac{A_j\,  \Gamma(j) \, \zeta(j+1)}{\beta^j} + D'(0)-D(0)\, \log \beta \right)\; (1+ \ord{\beta^{\epsilon}}).
\label{zapprox}
\end{equation}
as $\beta \to 0^+$.

We are now in a position to extract the density of states. First, observe that $\lZ_k(\beta)$ is actually a generating function for
\be
c_k(n) \equiv \sum_{ \bn \geq 0} \delta_{n_1+n_2+\cdots +n_k, n} \; d_k(\bn)
\ee
as can be seem by setting $\beta_i=\beta$ in (\ref{degdef}). Therefore we will derive an asymptotic formula for $c_k(n)$ as $n \to \infty$.
To derive this we use the integral expression \req{dosN}
\begin{equation}
c_k(n)= \frac{1}{2\pi i}\, \int_{b-i\pi}^{b+i \pi } \; d\beta\, \lZ_k(\beta)\, e^{n\,\beta}
\label{largedn}
\end{equation}
valid for any $b>0$. To obtain the saddle point evaluation of \req{largedn} consider
\begin{equation}
G_k(\beta)= \sum_{j=1}^k \frac{C_j}{j\, \beta^j} +n\, \beta  \ .
\label{Fdef}
\end{equation}
Then, for $\beta \to 0^+$  (\ref{zapprox}) tells us that $e^{G_k(\beta)}$ provides a good approximation to the integrand in \req{largedn}, with
\be C_j =  A_j \, \Gamma(j+1) \, \zeta(j+1). \ee We will choose the integration contour to pass through the saddle point of $G_k(\beta)$ with largest value of $G_k$, along the direction of the steepest descent. Let this extremum be $\beta_n$, so $G_k'(\beta_n)=0$, and thus we set $b=\beta_n$. We have
\begin{equation}
\sum_{j=1}^k \,\frac{C_j}{\beta_n^{j+1}}=n
\label{beta0def}
\end{equation}
and thus $\beta_n^{-1}$ is the largest positive root of this polynomial of order $k+1$. This shows that $n \to \infty$ is equivalent to $\beta_n \to 0$. More precisely
\begin{equation}
\beta_n =\left(\frac{C_k}{n}\right)^{\frac{1}{k+1}}\, \left(1+\ord{n^{-\frac{1}{k+1}}}\right)
\label{leadbeta0}
\end{equation}
and it is useful to record
\begin{equation}
G_k''(\beta_n)= \sum_{j=1}^k \frac{(j+1)\,C_j}{\beta_n^{j+2}} = \frac{(k+1)\,n^{\frac{k+2}{k+1}}}{C_k^{\frac{1}{k+1}}}  \left( 1+ \ord{n^{-\frac{1}{k+1}}} \right).
\label{Fderv}
\end{equation}
Note that since $G_k''(\beta_n)>0$ the direction of steepest descent is in the imaginary direction. Thus, now we change variables to $i\tau = \sqrt{G_k''(\beta_n)} \, \left(\beta-\beta_n\right)$. This implies that as $n\to \infty$ (for fixed $\tau$) we have\footnote{To derive this notice that $G_k^{(p)}(\beta_n) = \ord{\beta_n^{-p-k}}$. }
\be
G_k(\beta)= G_k(\beta_n) - \frac{1}{2}\tau^2+ \mathcal{O}(\beta_n^{k/2}).
\ee
It is also useful to note that $\log\beta= \log \beta_n+ \ord{\beta_n^{k/2}}$. Putting all this together and  we obtain
\begin{eqnarray}
c_k(n) &=&\, \int_{-\pi\,\sqrt{G_k''(\beta_n)}}^{\pi\,\sqrt{G_k''(\beta_n)}} \;  \frac{d\tau \, e^{D'(0)}}{2\pi\sqrt{G_k''(\beta_n)}}\; \exp\left(G_k(\beta_n) -D(0)\log \beta_n- \frac{1}{2}\tau^2 + \ord{\beta_n^{k/2}} \right)
 \times  \left(1+ \ord{\beta_n^{\epsilon}}\right)   \nonumber \\
&=& \sqrt{\frac{1}{2\pi G_k''(\beta_n)}}\,  \beta_n^{-D(0)} \; e^{G_k(\beta_n) +D'(0)} \; \left(1+ \ord{\beta_n^{\epsilon}}\right)\nonumber \\
&=& \frac{C_k^{\frac{(1-2D(0))}{2(k+1)}}}{\sqrt{2\pi (k+1)}} \;n^{\frac{2D(0)-k-2}{2(k+1)}} \;e^{G_k(\beta_n)+D'(0)} \left(1+ \ord{n^{-\frac{\epsilon}{k+1}}}\right)
\label{asymdn}
\end{eqnarray}
where, in order to perform the integral, the limits of $\tau$ have been replaced\footnote{The error in doing this is exponentially suppressed since $\pi\,\sqrt{G_k''(\beta_n)} \to \infty$  in the limit we are working.} by  $\pm\infty$. The last equality follows from using (\ref{leadbeta0}) and (\ref{Fderv}); note that the pre-factor multiplying the exponential is the same as Meinardus' result \cite{Meinardus:1953fk} as it comes from the rightmost pole. This is the desired asymptotic formula. Note, as we will show shortly, that we can obtain $\beta_n$ and $G_k(\beta_n)$ as a function of $n$ to the desired order using the defining polynomial for $\beta_n$ -- however, only the positive powers of $n$ in the expansion of $G_k(\beta_n)$ are meaningful in our asymptotic formula (and there are only a finite number of these). Also note that to leading order
\be
\log c_k(n) \sim G_k(\beta_n) \sim \left(1+ \frac{1}{k} \right)\, n^{\frac{k}{k+1}}\, \left[k\,\zeta(k+1)\right]^{\frac{1}{k+1}}
\label{singvardn}
\ee
which agrees with~\cite{Dolan:2007rq}.

Finally, observe that the above derivation is valid for any
Dirichlet series $D(s)$ with $k$ simple poles which converges for
$\Re{s}>k$ with residues $A_i$ -- in fact the number of poles and
the region of convergence need not be related.\footnote{One can take a  Dirichlet series which converges for $\Re{s} > k$ and has an arbitrary number of poles in the region $0< \Re{s} < k$; each of these poles then would contribute to the integrand in the saddle point evaluation.} Thus specialising our
formula to the case of a single simple pole at $s=k$, one can check
that it correctly reduces to Meinardus' theorem.

\paragraph{Explicit formulae:} We have shown how the asymptotics of $c_k(n)$ may be determined up to some known function $G_k(\beta_n)$. We will now address the explicit calculation of $G_k(\beta_n)$. First consider $k=2$. In this case $\beta_n^{-1}$ satisfies a cubic and thus we can get an analytic expression for it in terms of $n$. Explicitly we have
\be
\beta_n^{-3} + \frac{C_1}{C_2} \beta_n^{-2}-\frac{n}{C_2}=0
\ee
We want the largest positive root $\beta^{-1}_n$ to the above equation. Solving the cubic gives
\be
\beta_n^{-1}=-\frac{C_1}{3C_2} + y_n +\left(\frac{C_1}{3C_2} \right)^2 \frac{1}{y_n}
\ee
where
\be
y_n= \left( \frac{n}{2C_2}- \left(\frac{C_1}{3C_2} \right)^3 + \sqrt{ \left(\frac{n}{2C_2}- \left(\frac{C_1}{3C_2} \right)^3 \right)^2-\left(\frac{C_1}{3C_2} \right)^6} \right)^{1/3}.
\ee
From this it is a simple matter of computing the expansions of $\beta_n$, $G(\beta_n)$ for large $n$ which gives the explicit formula
\be
\label{explicit}
c_2(n) = \frac{C_2^{\frac{1}{6} - \frac{D(0)}{3}}}{\sqrt{6\pi}} n^{\frac{1}{3}(D(0)-2)} \exp\left( \frac{3}{2}C_2^{1/3} n^{2/3} + \frac{C_1}{C_2^{1/3}} n^{1/3} - \frac{C_1^2}{6\,C_2}+D'(0)\right) (1+ \ord{n^{-\epsilon/3}})
\ee
Observe that this illustrates the fact that for the leading asymptotics of $c_k(n)$ one needs more that just the contribution from the rightmost pole of $D(s)$. To see this, apply our formula (\ref{explicit}) to a partition function whose Dirichlet series has $C_1=0$ and call the density of states $\tilde{c}_2(n)$; it is then clear that $\lim_{n \to \infty} c_2(n)/\tilde{c}_2(n) \neq 1$, \ie, it is not true that $c_2(n) \sim \tilde{c}_2(n)$.

One can obtain explicit formulae for higher values of $k$ by a similar method. This is despite the fact there is no general formula for the root of a polynomial of order higher than 4. To get round this one can use the Lagrange inversion formula which tells one how to invert a Taylor series. Define $\lambda \equiv n^{-\frac{1}{k+1}}$ and $\beta_n \equiv y(\lambda)$, so $y \to 0$ as $\lambda \to 0$. From (\ref{beta0def})
\be
\lambda = \frac{y}{\phi(y)}, \qquad \phi(y) \equiv \left( \sum_{j=1}^k \,C_j\, y^{k-j} \right)^{\frac{1}{k+1}}.
\ee
The Lagrange inversion formula then gives
\be
y= \sum_{l=1}^{\infty} b_l \,\lambda^l, \qquad b_l \equiv \frac{1}{l!} \left[ \frac{d^{l-1}}{dy^{l-1}}\, \phi(y)^l \right]_{y=0}.
\ee
Notice that if $l$ is a multiple of $k+1$, so $l=(k+1)m$ for some integer $m$,  $\phi(y)^l$ is a polynomial of order $(k-1)m$ and thus $b_l=0$ (since $l-1= (k+1)m> (k-1)m$). Therefore
\be
\beta_n = \sum_{l>0, \; l \neq 0 \; \textrm{mod}\; (k+1)}^{\infty} b_l \; n^{-\frac{l}{k+1}}
\ee
from which it is a straightforward matter to extract the large $n$ expansion of $G(\beta_n)$ for any $k$. This shows how the computation of the positive powers of $n$ in the large $n$ expansion of $G(\beta_n)$ is purely algorithmic. To illustrate this we give the $k=3,4$ cases explicitly. For $k=3$:
\begin{eqnarray}
\beta_n &=& C_3^{\frac{1}{4}} \,n^{-\frac{1}{4}} + \frac{C_2}{4\,C_3^{\frac{1}{2}}} \,n^{-\frac{2}{4}} +\frac{1}{4 \,C_3^{\frac{1}{4}}} \left( C_1- \frac{C_2^2}{8\,C_3} \right) n^{-\frac{3}{4}} + \ord{n^{-\frac{5}{4}}}, \\
G_3(\beta_n) &=& \frac{4 \,C_3^{\frac{1}{4}}}{3} n^{\frac{3}{4}} + \frac{C_2}{2\,C_3^{\frac{1}{2}}} n^{\frac{1}{2}} + \frac{1}{C_3^{\frac{1}{4}}}\left( C_1- \frac{C_2^2}{8 \,C_3} \right) n^{\frac{1}{4}} \nonumber \\
& & \qquad  +\frac{C_2}{24 \,C_3}\left( \frac{C_2}{C_3}- 6C_1 \right)+ \;\ord{n^{-\frac{1}{4}}} \ ,
\end{eqnarray}
and for $k=4$ we get:
\bea
\beta_n &=& C_4^{\frac{1}{5}} n^{-\frac{1}{5}}+ \frac{C_3}{5\, C_4^{\frac{3}{5}}}n^{-\frac{3}{5}}+ \frac{1}{5\, C_4^{\frac{2}{5}}}\left( C_2- \frac{C_3^2}{5\,C_4} \right) n^{-\frac{3}{5}}\nonumber \\
& & \qquad + \frac{1}{5\,C_4^{\frac{1}{5}}} \left( C_1-\frac{C_2\,C_3}{5\,C_4} + \frac{C_3^2}{25 \,C_4^2} \right) + \ord{n^{-\frac{6}{5}}}, \\
G_4(\beta_n) &=& \frac{5\,C_4^{\frac{1}{5}}}{4} n^{\frac{4}{5}} + \frac{C_3}{3\,C_4^{\frac{3}{5}}} n^{\frac{3}{5}}+ \frac{1}{2\,C_4^{\frac{2}{5}}} \left( C_2- \frac{C_3^2}{5 \,C_4} \right)n^{\frac{2}{5}} +\frac{1}{C_4^{\frac{1}{5}}}\left( C_1- \frac{C_2\,C_3}{5\,C_4} + \frac{C_3^3}{25 \,C_4^2} \right)n^{\frac{1}{5}} \nonumber \\
&& \qquad +\frac{1}{5\,C_4}\left( -C_1C_3+ \frac{C_2\,C_3^2}{2\,C_4} - \frac{C_2^2}{2} - \frac{C_3^4}{12\, C_4^2} \right) + \ord{n^{-\frac{1}{5}}} \ .
\eea

\paragraph{Fermions:}
The partition function $\lZ^f_k(\beta)$ can be represented by an integral of the form (\ref{logZdir}) with $\zeta(s+1) \to \zeta_A(s+1)$ where $\zeta_A(s)=\sum_{n=1}^{\infty} (-1)^{n+1}/n^s$ is the alternating zeta function. Since $\zeta_A(s+1)=(1-2^{-s})\zeta(s+1)$ we see that $\zeta_A$ has no poles. Therefore the integrand of (\ref{logZdir}) will have a simple pole at $s=0$ (as opposed to a double one as in the bosonic case). This leads to the $\beta \to 0^+$ asymptotics being slightly modified:
\begin{equation}
\lZ^f_k(\beta) =\exp\left( \sum_{j=1}^k \; \frac{A_j\,  \Gamma(j) \, \zeta_A(j+1)}{\beta^j} + D(0)\log 2 \right)\; (1+ \ord{\beta^{\epsilon}}).
\label{zfapprox}
\end{equation}
and thus
\begin{equation}
c^f_k(n)= \sqrt{\frac{1}{2\pi \, {G_k}''(\beta_n)}}\; e^{G_k(\beta_n)+ D(0)\log 2} \; \left(1+\ord{\beta_n^{\epsilon}}\right)
\end{equation}
where $G_k(\beta)$ is given by \req{Fdef} with the coefficients $C_j = A_j \,\Gamma(j+1)\,\zeta_A(j+1)$. Again ${G_k}'(\beta_n)=0$ corresponds to the dominant saddle point. Observe that the leading order behaviour
\begin{equation}
\log c^f_k(n) \sim G_k(\beta_n) \sim \left(1+ \frac{1}{k} \right)\, n^{\frac{k}{k+1}}\;[k\, \zeta_A(k+1)]^{\frac{1}{k+1}}
\end{equation}
is the same (up to numerical factors) as in the bosonic case. Also note that one can use the explicit formulae developed above with the appropriate choice of $C_j$.

%_____________________________________
\subsection{Non-equal chemical potentials: multivariable generating functions}
\label{uneqchemp}
%_____________________________________
Now we will consider the asymptotics of the large $N$ generating functions  $\lZ_k(\bb)$, for non-equal chemical potentials $\beta_i$, as well as their associated density of states $d_k(\bn)$. It turns out that these generating functions can be written as products of the following functions:
\begin{equation}
F_k( \bb) \equiv \prod_{n_1,n_2, \cdots, n_k \geq 1}^{\infty} \, \frac{1}{1- \exp(- \bb \cdot \bn)}.
\label{}
\end{equation}
More precisely,
\be
\label{prodF}
\lZ_k(\bb)= \prod_{j=1}^k \; \prod_{i_1<i_2<\cdots<i_j} F_j(\beta_{i_1},\beta_{i_2},\cdots, \beta_{i_j}) \; .
\ee
 Thus it is convenient to concentrate on the $F_k(\bb)$ as they provide building blocks for the generating functions we are actually interested in.  Convergence of the infinite product is guaranteed by taking $\Re{\beta_i} > 0$. We will now derive the behaviour of this function as $\beta_i \to 0^++i0$. Of course this limit can be taken in a number of ways: we will be interested in when all the $\beta_i$ approach zero at the same rate, so $\beta_i= \gamma_i \beta$ with $\beta \to 0^+$ and $\gamma_i$ fixed. Taking the logarithm and expanding as before leads to
\begin{equation}
\log F_k(\bb) = \sum_{m,n_i \geq 1}\frac{ \,e^{-m \beta_in_i}}{m} = \frac{1}{2 \pi i} \int^{\gamma +i\infty}_{\gamma -i\infty} ds \; \Gamma(s) \,\zeta(s+1) \,\zeta(s; \bb)
\label{intrepF}
\end{equation}
where the second equality follows from using the inverse Mellin transform of the Gamma function (\ref{invMellin}) to replace the exponentials, and we have defined the following generalisation of the Riemann zeta function:
\begin{equation}
\zeta(s,\bb) \equiv \sum_{n_1, n_2, \cdots, n_k = 1}^{\infty} \,\frac{1}{(\bb \cdot \bn)^s}
\label{zetadef}
\end{equation}
which converges for $\Re{s} >k$ (so $\gamma > k$).

We will  need the full analytic structure of $\zeta(s, \bb)$ as a complex function of $s$. Fortunately this can be worked out in a manner analogous to the zeta function, or the Dirichlet series discussed in \sec{eqchemp}. This involves using the integral representation of the Gamma function to derive:
\begin{equation}
\zeta( s, \bb) = \frac{1}{\Gamma(s)} \int_0^{\infty} dt \; t^{s-1} \prod_{i=1}^k \frac{1}{e^{\beta_i t}-1}
\end{equation}
which again is only valid for $\Re{s}>k$ and then one can analytically continue this expression using the Hankel contour to get
\begin{equation}
\zeta( s, \bb) = -\frac{\Gamma(1-s)}{2\pi i} \int^{(0^+)}_{\infty}
dt \; (-t)^{s-1} \prod_{i=1}^k \frac{1}{e^{\beta_i t}-1} \equiv
-\Gamma(1-s) I(s,\bb).
\label{genzeta}
\end{equation}
Observe that $I(s,\bb)$ is an entire function of $s$. It vanishes
for $s=k+1, k+2, \cdots$. Therefore the apparent singularities
coming from $\Gamma(1-s)$ at these points are in fact removable and
therefore $\zeta(s, \bb)$ is analytic at these points. This leaves
$s=1,2 \cdots, k$ as potential poles, and indeed it is easy to see
$I(s,\bb)$ will be non-vanishing at these points, and thus they are
simple poles. Therefore the above expression provides the analytic
continuation of $\zeta(s, \bb)$ to the whole complex $s$ plane,
resulting in a meromorphic function with simple poles at $s=1, 2
\cdots, k$. The residues of these poles are
\begin{equation}
\textrm{Res}_{s=j} \; \zeta(s, \bb) = \frac{(-1)^{j+1}}{(j-1)!\;
\beta^j} \, I(j,\bgam)
\end{equation}
for $1 \leq j \leq k$ where
\begin{equation}
I(j,\bgam) = \textrm{Res}_{t=0} (-t)^{j-1} \,\prod_{i=1}^k
\,\frac{1}{e^{\gamma_i t}-1} = \frac{(-1)^{j-1}}{\prod_{i=1}^k
\gamma_i} \sum_{n_1,n_2, \cdots, n_k \geq 0} \delta_{n_1+ n_2+ 
 \cdots\, +n_k , k-j}\prod_{i=1}^k \frac{B_{n_i}}{n_i !} \gamma_i^{n_i} \;
\end{equation}
The second equality follows from using
$z/(e^z-1)=\sum_{n=0}^{\infty} B_nz^n/n!$ which defines the
Bernouilli numbers $B_n$. Note in particular, 
\begin{equation}
I(k,\bgam) =  \frac{(-1)^{k-1}}{\prod_{i=1}^k
\gamma_i} \ . 
\label{}
\end{equation}	

From the defining expression for $\zeta(s, \bb)$ it is easy to see that $\zeta(s,\bb)= \mathcal{O}(\beta^{-s})$ provided we avoid the poles. This implies that for all $0<\epsilon<1$
\begin{equation}
\int_{-\epsilon -i\infty}^{-\epsilon +i\infty} ds \;\Gamma(s)\, \zeta(s+1)\, \zeta(s, \bb) = \mathcal{O}( \beta^{\epsilon})
\label{error}
\end{equation}
as $\beta \to 0^+$, since the integrand has no poles in the $-1< \Re{s} < 0$ region of the complex $s$-plane. Therefore the asymptotic behaviour of $F_k(\bb)$ as $\beta \to 0^+$ is simply given by the sum of the residues of the poles of the integrand in (\ref{intrepF}) in the region $-\epsilon \leq \Re{ s} \leq k$ with the error in this estimate given by (\ref{error}).\footnote{Note that the contribution from the paths joining the contours $\Re{ s} =\gamma$ and $\Re{ s} =-\epsilon$ vanish due to the exponentially decaying behaviour of $\Gamma(s)$ in the imaginary direction.} The result of this analysis is:
\begin{eqnarray}
\label{asymptFk}
F_k(\bb) = \beta^{I(0, \bgam)}\; \exp \left[ \sum_{j=1}^k\,
(-1)^{j+1} \,\zeta(j+1) \,I(j,\bgam) \; \beta^{-j} +\zeta'(0,\bgam)
\right] \; \left( 1+ \mathcal{O}(\beta^{\epsilon}) \right)
\end{eqnarray}
as $\beta \to 0^+$. Note that to derive this we have used the following logarithmic property of our zeta function \req{genzeta}
\be
\label{log}
\zeta'(0,\bb)= \zeta'(0,\bgam)-I(0,\bgam)\log \beta \; .
\ee

Using these results it is a straightforward matter to deduce the
asymptotics of $\lZ_k(\bb)$ using (\ref{prodF}). For example, for $k=1$:
\begin{eqnarray}
\lZ_1(\beta) = F_1(\beta) = \sqrt{\frac{\beta}{{2\pi}} }\exp \left[ \frac{\pi^2}{6 \beta} \right] (1+ \mathcal{O}(\beta^{\epsilon}))
\end{eqnarray}
which is the well known result of Hardy and Ramanujan. The $k=2$ case is new and gives:
\begin{eqnarray}
\label{ktwo}
\lZ_2(\bb) &=& F_1(\beta_1)\, F_2(\beta_2) \, F_2(\bb) \\ \nonumber &=&
\frac{\beta^{\frac{1}{4}\left(3-
\frac{\gamma_1}{3\gamma_2}-\frac{\gamma_2}{3\gamma_1}\right)}}{2\pi}
\exp \left[ \frac{\zeta(3)}{\gamma_1\gamma_2 \beta^2} +\left(
\frac{\gamma_1+\gamma_2}{2\gamma_1\gamma_2} \right)
\frac{\pi^2}{6\beta} +\zeta'(0,\bgam)\right](1+
\mathcal{O}(\beta^{\epsilon}))
\end{eqnarray}
where $\bb = (\beta_1,\beta_2)=(\beta\gamma_1,\beta\gamma_2)$. 

For higher $k$ it is convenient to define the following zeta-function:
\be
\label{defZeta}
\Upsilon(s,\bb) \equiv \sum_{\bn > {\bf 0}} \, \frac{1}{\left(\bb.\bn\right)^s} = 
\sum_{j=1}^k \sum_{i_1<i_2<\cdots<i_j} \zeta(s,\beta_{i_1},\beta_{i_2},\cdots ,\beta_{i_j}) 
\ee
which also converges for $\Re{s} > k$. We may exploit the analytic continuation of $\zeta(s,\bb)$ to deduce that of $\Upsilon(s,\bb)$ in the complex $s$-plane, which is given by 
\begin{equation}
\Upsilon(s,\bb) = -\Gamma(1-s) \, \CI(s,\bb) \ ,
\label{}
\end{equation}	
where 
\begin{equation}
\CI(s,\bb) = \sum_{j=1}^k \sum_{i_1<i_2<...<i_j} \, I(s, \beta_{i_1},\beta_{i_2},\cdots, \beta_{i_j}) .
\label{calidef}
\end{equation}	
This shows that $\Upsilon(s,\bb)$ also  has simple poles at $s = 1, 2, \cdots , k$.\footnote{Of course, $\log \lZ_k(\bb)$ can be represented as integral (\ref{intrepF}) with $\zeta(s,\bb) \to \Upsilon(s,\bb)$, and we could have proceeded from here from the beginning. For clarity, though, we chose to analyse the $F_k(\bb)$ individually.} We can therefore write down the asymptotic formula for $\lZ_k(\bb)$ in a fashion analogous to \req{asymptFk} for $F_k(\bb)$; one has
\be
\label{asymptZk}
\lZ_k(\bb)=   \exp \left[ \sum_{j=1}^k\,
(-1)^{j+1} \,\zeta(j+1) \, \CI(j,\bb)  +\Upsilon'(0,\bb)
\right] \; \left( 1+ \mathcal{O}(\beta^{\epsilon}) \right)  .\ee
This asymptotic formula is one of the main results of this section. 

Now we are in a position to turn to the asymptotics of the density of states $d_k(\bn)$ of $\lZ_k(\bb)$. 
Thus we wish to evaluate
\begin{equation}
\label{dhat}
{d}_k(\bn) = \prod_{i=1}^k
\int_{b_i-i\pi}^{b_i+i\pi} \frac{d\beta_i}{2\pi i}\;
\lZ_k(\bb) \; e^{\bn \cdot \bb}
\end{equation}
in the limit $n_i \to \infty$ (at the same rate), which we will do by the saddle point method. The asymptotic formula (\ref{asymptZk}) tells us that $e^{G_k(\bb)}$, where
\begin{equation}
\label{Gbos}
G_k(\bb) \equiv  \bn \cdot \bb + \sum_{j=1}^k\,
(-1)^{j+1} \,\zeta(j+1) \, \CI(j,\bb)  \ ,
\end{equation}
provides a good approximation to the integrand for small $\beta_i$. We will choose the path of integration to go through a saddle point of $G_k$ defined by $\partial_i G_k=0$ such that $G_k$ is largest. We thus choose the $b_i$ appearing in (\ref{dhat}) to be this particular saddle point, which defines the $b_i$ as functions of the $n_i$. As we will see shortly, the Hessian $H_{ij}\equiv \partial_i\partial_j G_k$ evaluated at the dominant saddle point is positive definite (at least in the limit of interest), and therefore the path of steepest descent is in the imaginary direction. Now, observe that $H_{ij}(\bB)$ is a positive definite symmetric real matrix and thus must have real positive eigenvalues $\lambda^i$ such that $H_{ij}(\bB) e^l_j= \lambda^l e^l_i$ where $e^i_j$ is a set of orthonormal eigenvectors. This allows us to change variables from $\beta_i$ to $\tau_i$ defined by $\beta_i-b_i= i\tau_j\lambda_j^{-1/2} e^j_i$ so that
\begin{equation}
\label{Gtau}
G_k(\bb)=G_k(\bB)-\frac{1}{2} \tau_i\tau_i +\cdots
\end{equation}
where $\cdots$ denotes terms in the Taylor expansion of higher order. We can bound these higher orders terms as follows. Note that from $\partial_i G_k(\bB)=0$, which is a polynomial of total degree $k+1$ in the $1/b_i$, the limit $n_i \to \infty$ (all at the same rate) occurs when $b_i \to 0$ (at the same rate). In fact for small $\beta_i$
\be
G_k(\bb) \sim \bn \cdot \bb  + \frac{\zeta(k+1)}{\prod_{j=1}^k \beta_j}
\end{equation}
from which one can compute all derivatives $\partial_{i_1}\partial_{i_2}\cdots\partial_{i_p}G_k(\bB)$ in the limit $b_i \to 0$. Using $\partial_iG_k(\bB)=0$ this leads to:
\begin{equation}
b_i \sim \frac{ \zeta(k+1)^{\frac{1}{k+1}} \;\prod_{j=1}^k n_j^{\frac{1}{k+1}} }{n_i}, \qquad \prod_{j=1}^k b_j \sim \frac{\zeta(k+1)^{\frac{k}{k+1}}}{\prod_{j=1}^k n_j^{\frac{1}{k+1}}}
\label{asb}
\end{equation}
which shows explicitly that if we send all the $n_i \to \infty$ at the same rate then   $b_i \to 0$ at the same rate. Observe that the Hessian in this limit is
\be
H_{ij}(\bb) \sim \frac{\zeta(k+1)}{\prod_{j=1}^k \beta_j } \left( \frac{\delta_{ij}}{\beta_i^2} + \frac{1}{\beta_i \beta_j} \right) \ ,
\ee
from which it follows that
\begin{equation}
\textrm{det}[H_{ij}(\bb)]  \sim (k+1) \; \frac{\zeta(k+1)^k}{\left(\prod_{j=1}^k \beta_j \right)^{k+2}} \ .
\label{detG}
\end{equation}
Therefore $H_{ij}(\bB)$ is positive definite (at least in this limit) as promised earlier. We let $\bB=b\,{\bf c}$ where $b \to 0$ and ${\bf c}$ is fixed as $n_i \to \infty$; in particular, $b = \ord{|\bB|}$. Then it is easy to see that:
\begin{equation}
\partial_{i_1}\partial_{i_2}...\partial_{i_p}G_k(\bB)= \ord{|\bB|^{-k-p}}.
\end{equation}
Thus from $H_{ij}(\bB) = \ord{|\bB|^{-k-2}}$ we see that the eigenvalues $\lambda^j= \ord{|\bB|^{-k-2}}$ and hence for fixed $\tau_i$ we have $\beta_i-b_i= \ord{|\bB|^{(k+2)/2} }$, so
\begin{equation}
\partial_{i_1}\partial_{i_2}\cdots\partial_{i_p}G_k(\bB)\;(\beta_{i_1}-b_{i_1})\cdots(\beta_{i_p}-b_{i_p}) = \ord{ |\bB|^{(p-2)k/2}}
\end{equation}
and therefore for $p>2$ these terms are of order $\ord{|\bB|^{k/2}}$; this gives us a bound on the terms denoted by the ellipses in (\ref{Gtau}). One can also argue that for fixed $\tau_i$
\begin{equation}
\Upsilon'(0,\bb)=\Upsilon'(0,\bB)+  \mathcal{O}(|\bB|^{k/2})
\end{equation}
using (\ref{log}). Therefore we have established that for fixed $\tau_i$
\begin{equation}
 \label{key}
 \log
\lZ_k(\bb) + \bn \cdot \bb = G_k(\bB)- \frac{1}{2} \tau_i\tau_i+ \Upsilon'(0,\bB) + \ord{|\bB|^{\epsilon}}
\end{equation}
in the $n_i \to \infty$ limit which is of interest to us. Taking the Jacobian for the change of variables into account allows us to deduce
\begin{eqnarray}
\label{asymptdeg}
{d}_k(\bn) &=& \frac{1}{\sqrt{(2\pi)^k \det[ H_{ij}(\bB)]}} \exp\left( G_k(\bB)+ \Upsilon'(0,\bB) \right)\; (1+ \mathcal{O}(|\bB|^{k/2})) \\
& \sim & \left( \frac{\zeta(k+1)^{\frac{k}{k+1}}}{(2\pi)^k (k+1) \prod_{j=1}^k n_j^{\frac{k+2}{k+1}}} \right)^{1/2}\; \exp\left( G_k(\bB)+ \Upsilon'(0,\bB) \right)
\end{eqnarray}
where the second line follows from using (\ref{detG}) and (\ref{asb}). This asymptotic formula is the main result of this section. Note that to leading order
\begin{equation}
\label{lead}
\log {d}_k(\bn) \sim G_k(\bB) \sim (k+1) \, \zeta(k+1)^{\frac{1}{k+1}} \;\prod_{j=1}^k \, n_j^{\frac{1}{k+1}}
\end{equation}
which agrees with the result obtained in~\cite{Dolan:2007rq}. Also note that in the $k=1$ case (\ref{asymptdeg}) simplifies to the classic result of Hardy and Ramanujan (\ref{HR}).

%____________________________________
\subsubsection{Generalizations}
\label{genuneq}
%_____________________________________

\paragraph{Finite $N$:}
Now a word on the finite $N$ partition functions $\nZ_k(\bb,N)$. The $k=1$ case  was derived in \cite{Haselgrove:1954ax} (see \cite{Feng:2007ur} for application to superconformal theories), and our discussion will include this as a special case. There are a number of possible methods one can
contemplate to extract the density of states $d_{k,N}(\bn)$.  Logically, the most straightforward would be to determine the small
$\beta_i$ asymptotics of $\nZ_k(\bb,N)$, and then use the saddle point
method to deduce the asymptotic density of states, just as we did
above in the large $N$ case. However, the method we used above to
calculate the small $\beta_i$ asymptotics does not obviously lend
itself to $\nZ_k(\bb , N)$ as we do not have an (infinite) product
representation of it, the best we could do was write it in terms of
Bell polynomials (\ref{finNkpart}). A more indirect method, would be
to calculate the small $\beta_i$ asymptotics of $\Xi_k(\bb,p)$, for
fixed $p$, and use this to deduce the small $\beta_i$ asymptotics of
$\nZ(\bb , N)$ using
\be \label{intp}
\nZ_k(\bb , N) = \frac{1}{N!} \left(\frac{d^N}{dp^N}\;  \Xi_k(\bb,p) \right)_{p=0}. \ee
The advantage of
this, is that the method applied to the large $N$ case $\lZ_k(\bb)$ generalises
straightforwardly to $\Xi_k(\bb,p)$. In fact, following what we did
in the large $N$ case, note that
\be
\label{prodFp}
\Xi_k(\bb,p)= \frac{1}{1-p}\prod_{j=1}^k \prod_{i_1<i_2<\cdots<i_j} F_j(\beta_{i_1},\beta_{i_2},\cdots,\beta_{i_j},p)
\ee
where
\begin{equation}
F_k( \bb,p)= \prod_{n_1,n_2 \cdots n_k \geq 1}^{\infty} \, \frac{1}{1-p \,\exp(- \bb \cdot \bn)}
\label{fkpdef}
\end{equation}
which can be written as:
\begin{equation}
\log F_k(\bb,p) = \sum_{m,n_i \geq 1}\frac{ \,e^{-m \beta_in_i}}{m} = \frac{1}{2 \pi i} \int^{\gamma +i\infty}_{\gamma -i\infty} ds \; \Gamma(s) \,\textrm{Li}_{s+1}(p)\,\zeta(s; \bb).
\label{intrepFp}
\end{equation}
Note that this is of the same form as for $\log F_k(\bb)$
(\ref{intrepF}), with $\zeta(s+1) \to  \textrm{Li}_{s+1}(p)$. Recall
that $\zeta(s+1)$ has a simple pole at $s=0$. In contrast, the
polylogarithm function $\textrm{Li}_{s+1}(p)$ is an entire function
of $s$ for $|p|<1$ (since the defining sum converges absolutely for
these ranges). Therefore the pole structure of the integrand in
(\ref{intrepFp}) is simpler than in (\ref{intrepF}); the only change
being one has a simple rather than double pole at $s=0$. One can go
on to derive an analogous asymptotic formula for $F_k(\bb, p)$, which gives
\be
\label{finNasymp}
 F_k(\bb, p) = \exp\left[\sum_{j=0}^k \; (-1)^{j+1}\; \Li_{j+1}(p)\,  I(j,\bgam) \, \beta^{-j}
 \right]  \, \left(1+ \ord{\beta^\epsilon}\right) \;.
\ee
as $\beta \to 0^+$. This can then be used to deduce the asymptotics of $\Xi_k(\bb,p)$ using (\ref{prodFp}). Now we turn to the evaluation of $d_{k,N}(\bn)$ given by (\ref{dosN}). Using (\ref{intp}) and {\it assuming} we can swap the order of the $\beta_i$ integrations and the $p$-differentiations allows one to deduce the asymptotics of $d_{k,N}(\bn)$ for large $\bn$ via saddle point integration in $\beta_i$ as before resulting in:
\be
\label{asymptdkN}
d_{k,N}(\bn) \sim \frac{1}{N!}\, \frac{d^N}{dp^N}\left(\left( \frac{\Li_{k+1}(p)^{\frac{k}{k+1}}}{(2\pi)^k (k+1)(1-p)^2 \prod_{j=1}^k n_j^{\frac{k+2}{k+1}}} \right)^{1/2} \exp[G_k(\bB,p) ] \right)_{p=0}
\ee
where 
\begin{equation}
G_k(\bb,p) = \bn \cdot \bb +  \sum_{j=0}^k\; (-1)^{j+1}\; \Li_{j+1}(p)\, \CI(j,\bb) \; .
\end{equation}
As before $\bB$ is given by the location of the dominant saddle point $\partial_i G_k(\bb,p)=0$ and thus depends on $p$ as well as $\bn$; to leading order $\bB$ is given by (\ref{asb}) with $\zeta(k+1) \to \Li_{k+1}(p)$. It would be interesting to determine whether the steps leading to (\ref{asymptdkN}) are indeed valid and if so determine more explicitly its $N$ dependence.

\paragraph{Fermions:}  Let us mention what happens in the case of a fermionic
partition function. Thus define \be F^f_k(\bb)= \prod_{n_1,
n_2,...n_k \geq 1} 1+\exp(-\bb \cdot \bn) \ee which can then be
shown to satisfy (\ref{intrepF}) with $\zeta(s+1) \to
\zeta_A(s+1)\equiv (1-2^{-s}) \zeta(s+1)$. Therefore the pole at
$s=0$ is simple in this case and one gets
\begin{eqnarray}
F^f_k(\bb)= \exp \left[ \sum_{j=0}^k\,
(-1)^{j+1} \,\zeta_A(j+1) \,I(j,\bgam) \; \beta^{-j} \right] \left( 1+ \mathcal{O}(\beta^{\epsilon}) \right)
\label{ferFasy}
\end{eqnarray}
as $\beta \to 0^+$. This allows one to derive
\begin{equation}
 d^f_k(\bn) \sim \left( \frac{\zeta_A(k+1)^{\frac{k}{k+1}}}{(2\pi)^k (k+1) \prod_{j=1}^k n_j^{\frac{k+2}{k+1}}} \right)^{1/2} \exp[G_k^f(\bB) ] 
\label{ferasym}
\end{equation}
where
\begin{equation}
G_k^f(\bb) = \bn \cdot \bb +  \sum_{j=0}^k\; (-1)^{j+1}\; \zeta_A(j+1)\,  \CI(j,\bb)
\end{equation}
 and the dominant saddle $\partial_i G_k^f(\bB)=0$ defines $\bB$ (as a function of $\bn$), whose leading asymptotics are given by (\ref{asb}) with $\zeta(k+1) \to \zeta_A(k+1)$. As in the bosonic case $G_k^f(\bB)=\ord{|\bB|^{-k}}$ as $b \to 0^+$,  and hence
\begin{equation}
 \log {d}^f_k(\bn)
\sim G^f_k(\bB) \sim (k+1) \,\zeta_A(k+1)^{\frac{1}{k+1}} \;\prod_{j=1}^k
n_j^{\frac{1}{k+1}}
\end{equation}
and thus to leading order behaves in a
similar way to the bosonic partition functions. 

\paragraph{Weighted:}  In this case define
\be F^w_k(\bb)= \prod_{n_1,
n_2,\cdots,n_k \geq 1} \frac{1}{(1-\exp(-\bb \cdot \bn))^{w_{\bn}}} \ee
which can be shown to satisfy (\ref{intrepF}) with $\zeta(s,\bb) \to \zeta^w(s,\bb)$ where
\be
\zeta^w(s,\bb) = \sum_{n_1,n_2,\cdots,n_k \geq 1} \frac{w_{\bn}}{(\bb \cdot \bn)^s}.
\ee
and we will also consider $\Upsilon^w(s,\bb)$ defined as in (\ref{defZeta}) with $\zeta(s,\bb) \to \zeta^w(s,\bb)$.
The domain of convergence of these generalised Dirichlet series depends on $w_{\bn}$; let us suppose $\zeta^w(s,\bb)$ converges for $\Re s > \alpha$. One can represent this by an integral, valid for $\Re s >\alpha$ as we did in the simpler cases
\be
\zeta^w(s,\bb) = \frac{1}{\Gamma(s)} \int_0^\infty dt \; t^{s-1} \,g^w(t), \qquad g^w(t)= \sum_{ n_1,n_2,\cdots, n_k \geq 1} w_{\bn} \,e^{-\bb \cdot \bn \; t}.
\ee
The analytic continuation can then be performed using the Hankel contour:
\begin{equation}
\zeta^w(s,\bb) = -\Gamma(1-s)\,I^w(s,\bb) , \qquad I^w(s,\bb) \equiv \frac{1}{2\pi i}\, \int_{\infty}^{(0^+)} \, dt \, (-t)^{s-1}\,  g^w(t).
\end{equation}
The contour integral $I^w(s,\bb)$ defines an entire function of $s$. Thus the poles of $\zeta^w(s,\bb)$ must come from the Gamma function, and hence can only occur at the positive integers. Exactly which integers will depend on the analytic structure of $g^w(t)$ as $t=0$. If $g^w(t)$ has a pole of order $\alpha$, which is what we will assume, then $\zeta^w(s,\bb)$ will have at most\footnote{We say ``at most'' as it could occur that $I^w(s,\bb)$ vanishes at some of these points thereby removing the singularity.} simple poles at $s=1,2,\cdots,\alpha$ since $I^w(s,\bb)=0$ for all integer $s$ greater than $\alpha$. Then one can repeat the calculations to get the asymptotics of $F^w_k(\bb)$ as $\beta \to 0^+$:
\begin{eqnarray}
F^w_k(\bb) = \beta^{I^w(0, \bgam)}\exp \left[ \sum_{j=1}^{\alpha}\,
(-1)^{j+1} \,\zeta(j+1) \,I^w(j,\bgam) \; \beta^{-j} +{\zeta^w}'(0,\bgam)
\right] \left( 1+ \mathcal{O}(\beta^{\epsilon}) \right).
\end{eqnarray}
which then can be used to get the asymptotics of $\lZ_k^w(\bb)$ in the same way as (\ref{prodF}). From this one can derive:
\begin{equation}
\label{asymptdeg}
{d}^w_k(\bn) = \frac{1}{\sqrt{(2\pi)^k \det[ H^w_{ij}(\bB)]}} \exp\left( G_k^w(\bB)+ {\Upsilon^w}'(0,\bB) \right)\; (1+ \mathcal{O}(|\bB|^{\epsilon}))
\end{equation}
as $\bn \to \infty$, where $G_k^w(\bb)$ are as in (\ref{Gbos}) with $\CI(j,\bb) \to \CI^w(j,\bb)$, and  $\partial_i G^w_k(\bB)=0$ defines $\bB$ (as a function of $\bn$). Note that $\CI^w(s,\bb)$ is the weighted generalization of \req{calidef}. The leading asymptotic behaviour of $\log {d}^w_k(\bn)$ is given by $G^w_k(\bB)$, which in turn will be determined by the $j=\alpha$ term. However, given in general one does not know the exact form of $I^w(\alpha,\bb)$ (and thus $\bB$), we cannot be any more explicit at this stage without specifying $w_{\bn}$. Of course, given a specific $w_{\bn}$ one can easily apply the above formalism to get explicit answers.

\paragraph{Other asymptotics:}
We have only focused on asymptotics where the $n_i$ are all large and comparable. It is interesting to ask what occurs if one sends only some of the $n_i$ to infinity and keep the others fixed. It would then appear that one cannot use the saddle point method for the integrations in $\beta_i$ directions which correspond to the fixed $n_i$. Thus, at best, one can only perform a saddle point integration in the remaining directions, which require the knowledge of the $\beta_i \to 0^+$ asymptotics of the generating function in those directions alone (\ie, the other $\beta_i$ fixed). We will now outline how these may be worked out using the same techniques as before. The crucial step is to derive a different integral representation for $\log F_k(\bb)$ which is adapted to the asymptotics we require. Thus suppose we want the asymptotics as $\beta_i \to 0^+$ for $i=1,\cdots, p$ while keeping $\beta_i$ for $i=p+1,\cdots, k$ fixed. Expand $\log F_k(\bb)$ as in (\ref{intrepF}) and use the inverse Mellin transform of the Gamma function to replace the exponentials $e^{-n_i \beta_i m}$ only for $i=1,\cdots, p$. One can then resum the remaining $e^{-n_i \beta_i m}$ for $i=p+1,\cdots, k$ obtaining:
\be
\log F_k(\bb) = \int_{\gamma-i\infty}^{\gamma+i\infty} \frac{ds}{2\pi i} \; \Gamma(s)\; L(s+1;\hat{\bb}) \; \zeta(s, \bb')
\ee
where we denote $\bb'=(\beta_1,\cdots, \beta_p)$ and $\hat{\bb}= (\beta_{p+1},\cdots,\beta_k)$ and we have defined the function
\be
L(s;\hat{\bb}) = \sum_{m=1} \frac{l_m(\hat{\bb})}{m^{s}}, \qquad l_m(\hat{\bb})= \prod_{j=p+1}^k \frac{1}{e^{\beta_j m}-1} \; .
\ee
Note that, just as for $\Li_s(p)$ for $|p|<1$, the series defining $L(s;\hat{\bb})$ converges absolutely for all complex $s$ and thus defines an entire function in the complex $s$-plane. Therefore, as before, we may work out the asymptotics for small $\bb'$ resulting in:
\be
F_k(\bb) = \exp \left[ \sum_{j=1}^p (-1)^{j+1}\; L(j+1;\hat{\bb})\; I(j,\bgam')\; \beta^{-j} \right] (1+ \ord{\beta^{\epsilon}})
\ee
where $\bb'=\beta \bgam'$ and $\beta \to 0^+$. From this, using (\ref{prodF}), one can deduce the asymptotics for $\lZ_k(\bb)$ itself in this particular limit. Now let us turn to the question of interest: the asymptotics of $d_k(\bn)$ in the limit of large $\bn'=(n_1,\cdots,n_p)$ with $\hat{\bn}=(n_{p+1},\cdots,n_k)$ fixed. Note that one can write:
\be
d_k(\bn)= \prod_{j=p+1}^k \frac{1}{n_j!} \frac{d^{n_j}}{dx_j^{n_j}}  \;\left( \prod_{i=1}^p
\int_{b_i-i\pi}^{b_i+i\pi} \frac{d\beta_i}{2\pi i}\;
\lZ_k(\bb) \; e^{\bn' \cdot \bb'} \; \right)_{x_j=0}
\ee
where $x_j=e^{-\beta_j}$.
If one {\it assumes} that the large $\bn'$ limit and the $x_j$ differentiations can be swapped, then one can perform a saddle point integration in the $\bb'$ direction (as this localises around $\bb'=0$); this then requires the asymptotics of $\lZ_k(\bb)$ for small $\bb'$ which we have worked out above. Carrying out such steps gives the required asymptotics of $d_k(\bn)$ in terms of $x_j$ derivatives which we will not write down. It would be interesting to determine whether this procedure is in fact valid.

%~~~~~~~~~~~~~~~~~~~~~~~~~~~~~~~~~~~~~~~~~~~~~~~
\section{Limit curves and typical operators}
\label{limcr}
%~~~~~~~~~~~~~~~~~~~~~~~~~~~~~~~~~~~~~~~~~~~~~~~
Let us now consider the partition functions $\nZ_k(\bb,N)$ and $\lZ_k(\bb)$ from a more statistical point of view. We would like to infer from these partition sums the class of ``typical operators" which dominate the ensemble under consideration. In general from all the states in the Hilbert space $\CH_{BPS}$ we will find that one can identify a limit curve in the charge space on which the typical operators lie. Our analysis follows the statistical treatment of \cite{Vershik:1996uq,Vershik:1996fk}, which has previously been used in the discussion of $\frac{1}{2}-$BPS states in $\CN=4$ SYM in \cite{Balasubramanian:2005mg} and more recently for quiver $\CN=1$ theories in \cite{Balasubramanian:2007hu}.

To begin with let us introduce a probability distribution on the set of quantum numbers $\bn$ by:
\begin{equation}
p_k(\bn,N)= \frac{d_N(\bn)\, e^{-\bb \cdot \bn}}{\nZ_k(\bb,N)}
\end{equation}
so $\sum_{ \bn>0} p_k(\bn,N)=1$. We may now define the expectation value and variance of $\bn$ in a standard fashion
\begin{eqnarray}
\langle n_i \rangle &=& - \frac{ \partial}{\partial \beta_i} \log \nZ_k(\bb,N)  \\
\textrm{var}[n_i] & \equiv & \langle n_i^2 \rangle- \langle n_i \rangle^2 = \frac{ \partial^2}{\partial \beta_i^2} \log \nZ_k(\bb,N).
\end{eqnarray}
We will be now interested in features of this distribution in the limit of large $\langle n_i \rangle$ all of the same order.
Physically the $\beta_i$ have the interpretation of the inverse temperature of a canonical ensemble and thus we expect the only way to achieve $\langle n_i \rangle \to \infty$ is to send $\beta_i \to 0$ (at the same rate). However this temperature does not have any meaning intrinsic to the theory which the states belong to, it simply sets the mean ``energy''.
In any case, as $\beta_i \to 0$ we can use our leading order result for $\nZ_k(\bb,N)$ (\ref{finiteNasymp}) to obtain
\begin{eqnarray}
\langle n_i \rangle \sim \frac{N}{\beta_i}, \qquad  \textrm{var}[n_i]  \sim \frac{N}{\beta_i^2}
\end{eqnarray}
and therefore
\begin{equation}
\frac{\sqrt{ \textrm{var}[n_i] }}{\langle n_i \rangle} \sim \frac{1}{\sqrt{N}}.
\end{equation}
Note that these expressions are valid for all $N$ in the $\beta_i \to 0$ limit. Therefore we see that in the large $N$ limit these $\beta_i  \to 0$ distributions are sharply peaked around the mean. This allows one to introduce the concept of a limit curve \cite{Vershik:1996uq,Vershik:1996fk} which we now discuss. Let us work with the strict large $N$ limit partition function $\lZ_k(\bb)$ since as we have just argued this is the regime where the distributions localise around the mean the most. One finds the exact answers
\begin{eqnarray}
\langle n_i \rangle &=& \sum_{\bn >0} \frac{n_i}{e^{\bb \cdot
\bn}-1} \sim \frac{\zeta(k+1)}{\beta_i \prod_{j=1}^k \beta_j}, \\
\textrm{var}[n_i] &=& \sum_{\bn >0} \frac{n_i^2}{4 \sinh^2 \left(
\frac{\bb \cdot \bn}{2}\right) } \sim \frac{2 \zeta(k+1)}{\beta_i^2
\prod_{j=1}^k \beta_j}.
\end{eqnarray}
where the asymptotics for $\beta_i \to 0$ can be deduce from our asymptotics for $F_k(\bb)$ \req{asymptFk}, noting that $\log \lZ_k(\bb) \sim \log F_k(\bb)$. It thus follows that
\begin{equation}
\frac{\sqrt{ \textrm{var}[n_i] }}{\langle n_i \rangle} \sim
\sqrt{\frac{2}{\zeta(k+1)}} \prod_{j=1}^k \beta_j^{1/2}
\end{equation}
which indeed tends to zero as predicted above from the large $N$ limit of the finite $N$ answer, although this also tells us how fast it tends to zero.
As just discussed in the $\beta_i \to 0$ limit these distributions localise around the mean and due to the combinatorial interpretation in terms of vector partitions we may read off the so called completion numbers of the ``partition'' of $\langle n_i \rangle$. For a vector partition of say $\bf{v}$, the completion numbers $r_{\bf{v}}(\bn)$ are defined by ${ \bf v}= \sum_{\bn>0} r_{\bf{v}}(\bn) \bn$, \ie, they give the number of times $\bn$ appears in the given partition of $\bf{v}$. Thus, we may formally define the {\it completion numbers} of $\langle n_i \rangle$, despite this not necessarily being an integer, by
\begin{equation}
\label{comp}
r_{\langle \bn \rangle}(\bn) = \frac{1}{e^{\bb \cdot \bn}-1}.
\end{equation}
Now, given a set of completion numbers one may define the following curve, or surface, in $\mathbb{R}^{k+1}$ given by $(t_1,, t_2, \cdots t_k, \phi({\bf t}) )$ where
\begin{equation}
\phi({\bf t}) = \sum_{\bn \geq {\bf t}} r_{\bf{v}}(\bn).
\end{equation}
Observe that $\int_{0}^{\infty} dt_i \; \phi_i(t_i) = v_i$ where we have defined $\phi_i(t_i)=\phi({\bf t})_{t_j=0, j \neq i}$. It is clear that $\phi({\bf t})$ is not a smooth, or even continuous function for vector partitions of $\bf{v}$. However, the curve one obtains using the ``completion numbers'' of $\langle n_i \rangle$ (\ref{comp}), is a smooth  function of ${\bf t}$. In the limit of interest,  $\langle n_i \rangle \to \infty$ with $\frac{\sqrt{ \textrm{var}[n_i] }}{\langle n_i \rangle} \to 0$,  one may replace the sum defining $\phi({\bf t})$ with an integral. To do this explicitly, since $\int_0^{\infty}dt_i \phi_i(t_i) = \langle n_i \rangle$, we first need to perform a rescaling in order that the limit may be taken. Thus, define
\begin{equation}
\tilde{\phi}({\bf t}) \equiv \frac{\prod_{j=1}^k
\beta_j}{\zeta(k+1)}\; \phi\left(\frac{t_1}{\beta_1}, \cdots ,\frac{t_k}{\beta_k}\right)
\end{equation}
so that
\begin{equation}
\int_0^{\infty} dt_i \; \tilde{\phi}_i(t_i) = \frac{\langle n_i
\rangle \beta_i \prod_{j=1}^k \beta_j}{ \zeta(k+1)} \to 1
\end{equation}
 as $\beta_{i} \to 0$.
Then, defining $y_i=\beta_i n_i$, we see that:
\begin{equation}
\tilde{\phi}({\bf t}) =  \frac{1}{\zeta(k+1)} \sum_{{\bf y} \geq
{\bf t}}\frac{\prod_{j=1}^k \beta_j}{e^{{\bf 1}\cdot {\bf y}}-1} \to
\frac{1}{ \zeta(k+1)}\int_{{\bf y \geq t}} \prod_{i=1}^k \,dy_i
\,\frac{1}{e^{{\bf 1} \cdot {\bf y}}-1} = \frac{1}{\zeta(k+1)}\,
\textrm{Li}_{k}( e^{-{\bf1} \cdot {\bf t}})
\end{equation}
which defines the limit curve $C({\bf t})$. Hence for the bosonic partition sum \req{lzkbos}
\begin{equation}
C({\bf t}) =  \frac{1}{\zeta(k+1)}\,
\textrm{Li}_{k}( e^{-{\bf1} \cdot {\bf t}}) .
\label{limcbk}
\end{equation}
Observe that
$\int_{0}^{\infty} dt_i \; C_i(t_i)=1$ which was guaranteed by the
construction.

\paragraph{Fermions:} For the applications we have in mind it is also useful to obtain the fermionic counterpart of the limit curve $C({\bf t})$ defined above. From the fermionic partition function
\be
\lZ_k^f(\bb) = \prod_{\bn \geq 0} 1+ e^{-\bb \cdot \bn} \ , \ee
one obtains
\begin{eqnarray}
\langle n_i \rangle &=& \sum_{\bn >0} \frac{n_i}{e^{\bb \cdot
\bn}+1} \sim \frac{\zeta_A(k+1)}{\beta_i \prod_{j=1}^k \beta_j}, \\
\textrm{var}[n_i] &=& \sum_{\bn >0} \frac{n_i^2}{4 \cosh^2 \left(
\frac{\bb \cdot \bn}{2}\right) } \sim \frac{2 \zeta_A(k+1)}{\beta_i^2
\prod_{j=1}^k \beta_j}.
\end{eqnarray}
where the asymptotics for $\beta_i \to 0^+$ are deduced from our asymptotics for $F^f_k(\bb)$ \req{ferFasy}, noting that $\log \lZ^f_k(\bb) \sim \log F^f_k(\bb)$. Then
\begin{equation}
\frac{\sqrt{ \textrm{var}[n_i] }}{\langle n_i \rangle} \sim
\sqrt{\frac{2}{\zeta_A(k+1)}} \prod_{j=1}^k \beta_j^{1/2}
\end{equation}
which shows that in the $\beta_i \to 0$ limit this distribution is sharply peaked around the mean as in the bosonic case. Therefore in this limit the concept of a limit curve is sensible and repeating the steps above
leads to the limit curve:
\be C^f_k({\bf t})= -\frac{1}{\zeta_A(k+1)}\,
\textrm{Li}_{k}( -e^{-1 \cdot {\bf t}}).
\label{limcfk}
\ee

\paragraph{Applications:} Having obtained the behaviour of the limit curves for the basic partition functions of interest, we now turn to some applications of this concept.
Observe that the bosonic $k=1$ case corresponds to a curve in
$\mathbb{R}^2$; one way to interpret this is to think of it as describing the behaviour of Young Tableaux's in the limit described above (recall that the standard integer partitions are naturally associated with Young Tableaux). In particular, the limit curve defined in \req{limcbk} for $k=1$ is basically the typical Young tableaux in the ensemble. This is related to the concept of ``typical" $\frac{1}{2}$-BPS operators in $\CN=4$ SYM \cite{Balasubramanian:2005mg}.\footnote{This limit curve is also relevant for thermal ensembles in a free boson theory in two dimensions \cite{Balasubramanian:2007qv}.} From the dual supergravity in \ads{5} $\times S^5$ description $\frac{1}{2}$-BPS states correspond to smooth geometries which are also specified by a curve in $\mathbb{R}^2$ (the LLM plane). Indeed, using the limit curve as a boundary condition for the supergravity solutions leads to the so-called ``hyperstar'' (a singular geometry) intepreted as the effective geometry dual to generic $\frac{1}{2}$-BPS operators of fixed charge \cite{Balasubramanian:2005mg}.

An interesting question is whether such a mapping occurs in the
$\quarter-$BPS and $\eighth$-BPS sectors. The $k=2$ bosonic case above gives the limit curve for $\quarter-$BPS operators and consists of a 2-surface in
$\mathbb{R}^3$. The supergravity point of view is far less developed than in
the $\frac{1}{2}$-BPS case, essentially due to the non-linearity of the
resulting equations -- see  \cite{Donos:2006iy} for the supergravity analysis. However, it has still been argued that smooth geometries should exist, specified by a smooth 3-surface in a 4d K\"ahler space \cite{Chen:2007du} (see also the recent analysis of \cite{Lunin:2008tf}). This therefore raises a puzzle as the boundary
conditions for these supergravity solutions consist of a surface of
different dimensionality to the limit curve.

The $\eighth$-BPS case as we have been earlier includes contributions from
fermionic partitions functions. In order to construct the limit
curve for this observe the following fact. Suppose we have two
partition functions $\lZ_1$ and $\lZ_2$ which generate some class of
vector partitions with associated limit curves $C_1$ and $C_2$. The
limit curve associated to $\lZ_1\,\lZ_2$ is $C_1+C_2$. Armed with this and
the above results we can try and compute the limit curve of the
large $N$ $\eighth$-BPS partition function (\ref{piecesofeight}). The
special case $e^{\zeta}= 1/{x_1x_2x_3}$ is straightforward as it
can be written in terms of our basic partition functions:
\be \lZ^{(\eighth)}(\bb,\gamma) =
\frac{\lZ_3(\bb)^2}{\prod_{i<j}\lZ_2(\beta_i,\beta_j)} \times
\frac{\lZ^f_3(\bb)^2}{\prod_{i<j}\lZ^f_2(\beta_i,\beta_j)}.\ee
Thus the limit curve in this case is:
\be C^{(1/8)}({\bf t}) = 2\, C_3({\bf
t})+2\, C^f_3({\bf t})- \sum_{i<j} \, \left[C_2(t_i,t_j) + C^f_2(t_i,t_j) \right] \; .\ee
It would be interesting to work out the limit curve for general
chemical potentials, but we shall not pursue this here. The $\eighth$-BPS limit curve above consists of a 3-surface in a $\mathbb{R}^4$ and presumably including the extra chemical potential would only increase its dimensionality by one. The supergravity situation in this sector is even less understood \cite{Gava:2006pu}, but it seems smooth solutions should be specified by 5d boundary conditions in some 6d K\"ahler space \cite{Chen:2007du} and thus again it is not clear how this relates to the limit curve we have just discussed.

%~~~~~~~~~~~~~~~~~~~~~~~~~~~~~~~~~~~~~~~~~~~~~~~
\section{Discussion}
\label{discuss}
%~~~~~~~~~~~~~~~~~~~~~~~~~~~~~~~~~~~~~~~~~~~~~~~

In this paper we have performed a detailed analysis of chiral ring partition functions for $D=4$ superconformal field theories (and analogous BPS states in $D=3,6$) . In particular, we have shown how the knowledge of the generating function for the finite $N$ partition function, aided with a suitable combinatorial interpretation, can be used to explicitly write down the finite $N$ partition functions. These capture the physics of the operators generating the chiral ring away from the planar limit. We have also presented detailed analysis of the density of states and discussed the notion of typical states for these field theories.

 The finite $N$ result derived above should be useful to understand the behaviour of $\frac{1}{4}$-BPS and bosonic $\frac{1}{8}$-BPS operators in $\CN =4$ SYM, whose finite $N$ partition functions are generated by the grand-canonical partition sum \req{zkdef} with $k=2$ and $k=3$ respectively \cite{Kinney:2005ej}.\footnote{Similar statements can be made for M2 and M5 brane world-volume theories \cite{Bhattacharyya:2007sa} and our considerations can easily be extended to these cases.} One motivation for this analysis was to investigate whether it is possible to come up with a
 simple auxiliary model to understand the dynamics of these operators. For instance for $\frac{1}{2}$-BPS operators one can use the fact that the system has a representation in terms of free fermions \cite{Berenstein:2004kk}, which has been exploited to understand the detailed behaviour of these states from a dual supergravity perspective \cite{Lin:2004nb}.

  A natural question is whether there is a similar simplification for $\frac{1}{4}$ and $\frac{1}{8}$-BPS operators?\footnote{See recent discussions in  \cite{Dutta:2007ws, Azuma:2007fj} where it is argued that perhaps even the non-supersymmetric states related to black hole geometries might enjoy a free fermion description.} This  has been discussed in \cite{Berenstein:2005aa} where the answer is argued to be in the affirmative. Focussing on the chiral ring, \cite{Berenstein:2005aa} postulates that the dynamics of $\frac{1}{8}$-BPS states can be encoded by an $SO(6)$ matrix model of commuting matrices (see \cite{Berenstein:2007wi,Berenstein:2007kq} for conjectured extensions to $\CN=1$ field theories). One check of this proposal is that it should be able to reproduce the correct spectrum of the states in question.  While it is possible in this matrix model to write down the ground state, the exact spectrum of excited states is not amenable to analytic computation, owing to the complicated eigenvalue interaction potential arising from the measure.\footnote{This is computed using a {\it flat} metric on the space of matrices and rewriting this in terms of the eigenvalues and the off-diagonal elements. Integrating out the off-diagonal elements leads to the desired measure factor.}  It has been previously argued in \cite{Agarwal:2006nv} that this particular matrix model cannot have a free fermion representation and they provide explicit analysis of the interaction terms in question. Our analysis is complementary and demonstrates from a spectral viewpoint that the system is not simply governed by free fermion dynamics. In particular, our factorisation result shows that the partition function for the finite $N$ $\quarter$-BPS chiral ring of $\CN=4$ SYM can be written as
\be
Z^{\frac{1}{4} \textrm{BPS}}_N(x_1,x_2)= Z^{\frac{1}{2} \textrm{BPS}}_N(x_1) \; Z^{\frac{1}{2} \textrm{BPS}}_N(x_2) \; P_N(x_1,x_2)
\ee
where $P_N(x_1,x_2)$ is a symmetric polynomial of order $\frac{1}{2}N(N-1)$ in each $x_i$ with non-negative coefficients. This very clearly illustrates that the $\quarter$-BPS operators can be thought of as consisting of two sets of different $\frac{1}{2}-$BPS operators together with some interaction governed by the polynomial.  It would be interesting to understand the implication of the factorisation result and use it to decode more detailed properties of the chiral ring in $\CN =4$ SYM.

 The main mathematical result of the paper is a derivation of the asymptotic density of states for the multi-variable partition functions in question. A special case of this includes a generalization of Meinardus' theorem to the case where the associated Dirichlet series has multiple poles. These generalizations provide a concrete algorithmic way to decipher the asymptotic growth of states in the superconformal theories. Consider the leading order estimate (\ref{lead}) -- for integers which scale as $N^2$ (\ie, corresponding to operators whose charges scale as $N^2$) this grows as $\log d(N^2) = \ord{ N^{\frac{2 k}{k+1} }}$.  In four dimensional field theories with holographic duals, operators of conformal dimension $\ord{N^2}$ are generically expected to be dual to heavy states such as black holes. However, since $\frac{2k}{k+1}<2$ the growth of the density of states in the chiral ring is not enough to account for black hole entropy (a fact noticed by many authors previously \cf, \cite{Kinney:2005ej}).  Not surprisingly, similar results hold for superconformal field theories in other dimensions; for example $\frac{1}{8}-$BPS  states of the M2-brane worldvolume theory has a growth of states given by  $\log d_{{\rm M2}}(n)= \ord{ n^{4/5}}$; here the states of interest should have conformal dimension $\ord{N^{3/2}}$ so obtain a non-trivial back-reaction in the dual AdS background. The density of such states grows at most as $N^{6/5} < N^{3/2}$. While this result can be inferred directly without recourse to Meinardus' theorem or generalizations thereof, our analysis provides a useful characterization of the sub-leading terms. Assuming one were to be able to construct explicit ``small-black hole'' solutions (\ie, gravitational solutions which incorporate higher derivative corrections) dual to these operators and reproduce the leading order growth of the density of states, one can then analyze the sub-leading corrections using our formalism.

Our analysis also touched upon the issue of typical operators in the supersymmetric sectors; the charge vectors for these operators  lie close to the limit curves we derived. It would be interesting to understand the relation between the limit curves and the class of typical smooth solutions in supergravity.

On a more technical side, there are a number of open problems. For instance, we have
briefly touched upon the asymptotics of the finite $N$ partition sums. While the technology we developed can be used to extract the asymptotics of the grand canonical partition sum, it would be useful to have more explicit finite $N$ asymptotic formulae. Further, we have   only discussed the asymptotics for the mesonic operators in $\CN =1 $ field theories. Generically, these theories also have baryonic operators and the partition sums receive contributions from non-zero baryon number sectors.  Another interesting class of partition sums are the ones that occur in the free theory -- these are typically expressed as matrix integrals \cite{Aharony:2003sx}. Like the baryonic partition sums and the finite $N$ results, these generically are not of infinite product form, thereby requiring new ideas to extract precise asymptotic formulae.   It would be interesting to develop the technology to determine the asymptotics of these more general partition sums.

%~~~~~~~~~~~~~~~~~~~~~~~~~~~~~~~~~~~~~~~~~~~~~~~
\subsection*{Acknowledgements}
\label{acks}
%~~~~~~~~~~~~~~~~~~~~~~~~~~~~~~~~~~~~~~~~~~~~~~~
It is a pleasure to thank Amihay Hanany and Shiraz Minwalla for discussions and especially Francis Dolan for extensive comments on a draft version of the paper. This work was supported by STFC.
%end
%%%%%%%%%%%%%%%%%%%%%%%%%%%%%%%%%%%%%%%%%%%%%
\bibliography{counting}
\bibliographystyle{utphys}

\providecommand{\href}[2]{#2}\begingroup\raggedright\endgroup

\end{document}